\documentclass[12pt,preprint]{article}

\usepackage[textwidth=20.4cm, textheight=25cm]{geometry}
\usepackage{amsmath}
\usepackage{amssymb}
\usepackage{graphicx}
\usepackage{hyperref}
\usepackage{cite}
\usepackage{authblk}
\usepackage{xcolor}

\title{Symplectic tomographic probability distribution
of crystallized Schr\"odinger cat states}
\author[1,2,3]{Julio A. L\'opez-Sald\'ivar}
\author[4]{Margarita A. Man'ko}
\author[2,3,4]{Vladimir I. Man'ko}

\affil[1]{Instituto de Ciencias Nucleares, Universidad Nacional Aut\'onoma de
M\'exico, Apdo. Postal 70-543, Ciudad de México 04510, M\'exico}
\affil[2]{Moscow Institute of Physics and Technology, Institutskii per. 9,
Dolgoprudnyi, Moscow~Region 141700, Russia} \affil[3]{Russian Quantum Center,
Skolkovo, Moscow 143025, Russia} \affil[4]{Lebedev Physical Institute,
Leninskii Prospect 53, Moscow 119991, Russia}
\date{\small julio.lopez.8303@gmail.com}

\begin{document}
\maketitle

\abstract{Within the framework of the probability representation of quantum
mechanics, we study a superposition of generic Gaussian states associated to
symmetries of a regular polygon of $n$ sides; in other words, the cyclic
groups (containing the rotational symmetries) and dihedral groups (containing
the rotational and inversion symmetries). We obtain the Wigner functions and
tomographic probability distributions (symplectic and optical tomograms)
determining the density matrices of the states explicitly as the sums of
Gaussian terms. The obtained Wigner functions demonstrate nonclassical
behavior, i.e., contain negative values, while the tomograms show a series of
maxima and minima different for each state, where the number of the critical
points reflects the order of the group defining the states. We discuss general
properties of such a generalization of normal probability distributions.}

\section{Introduction}
The states of quantum systems can described by different approaches: as wave
functions~\cite{Schroedinger26}, as density matrices~\cite{Landau,vonNeimann}, or using standard probability distributions called the tomographic representation of quantum states~\cite{ManciniPLA,IbortPS150,Olga-Entropy21}. In particular, 
even and odd superpositions of Gaussian wave functions of standard coherent
states (i.e., coherent states considered by Glauber~\cite{Glauber63} and Sudarshan~\cite{Sudar63}) have been very important for the development of quantum theory as they present non-classical behavior in spite of being the superposition of quasi-classical states. These states called even and odd Schr\"odinger cat states were studied
in~\cite{Sanders-PR,Sanders-JPA, DodPhysica1974}. They have specific symmetry properties
associated to the group of two elements -- the identity and mirror reflection operations in the phase space. The even and odd coherent
states were studied in the probability representation
in~\cite{Olga-SPIE,Adam}. The states with other symmetries, such as
crystallographic symmetries, were applied to Schr\"odinger cat states
in~\cite{manko2,manko3,castanos}. In~\cite{entan}, different ways to detect
the bipartite entanglement, using the tomographic probability representation,
were presented. Additionally, it has been established that the generalization of the symmetric superposition of coherent states can be used as a qudit system as they are orthogonal between each other \cite{generalstates}.

Several procedures have been proposed to obtain odd and even coherent states. For
example, there was a proposal to generate low-photon-number cat states, also
known as kitten states, in~\cite{ourjoumtsev1}. Other proposals have been made, namely, (i)~to reflect a coherent pulse in an optical cavity with one
atom~\cite{hacker,wang}; (ii)~to detect a photon-number
state~\cite{ourjoumtsev2} employing  the ancilla-assisted photon
subtraction~\cite{takahashi}; (iii)~to subtract a particular number of photons
from a squeezed vacuum state~\cite{gerrits} or to subtract one photon of a
squeezed vacuum state~\cite{neegaard}. Nonclassical features of the
superposition of coherent states like squeezing have been studied
in~\cite{buzek,janszky,domokos}. The possible experimental implementation of
these superpositions was mentioned in~\cite{szabo}, and this implementation
for coherent states on a circle was discussed
in~\cite{janszky,domokos,gonzalez}. The circle states are connected with the
phase--time operators in the problem of harmonic
oscillator~\cite{susskind,nieto1,pegg1,pegg2}. States with circle symmetry
were defined in~\cite{calixto}, using spin coherent states; also
in~\cite{calixto1}, the $su(1,1)$ coherent states on a hyperboloid were
explored. A proposed method to generate states with high symmetry was studied
by taking into account the dynamic evolution of a matter--field interaction
described by the Tavis--Cummings Hamiltonian~\cite{cordero1,cordero2}.

The particular states studied in this work are associated to the symmetries of the regular $n$-sided polygon, in other words, superposition states resulting from the cyclic and dihedral group of operations over an initial state. The cyclic group $C_n=\mathbb{Z}/(n \mathbb{Z})$ is an abelian group, which
elements are the operations associated to the rotation symmetries for a
regular polygon. In other words, the $n$-degree cyclic group can be defined as
$C_n=\{I, R(2\pi/n), R(4\pi/n), \ldots , R(2(n-1)\pi/n) \}$, where $R(\theta)$
is the rotation operation. The cyclic group $C_n$ has $n$ irreducible
representations denoted by $\lambda=1,\ldots, n$. The character
$\chi_r^{(\lambda)}$ associated to the $r$th element of the group for the
irreducible representation $\lambda$ is given by one of the roots of unity.

The dihedral group $D_n$ is a non-abelian group, with elements being the
rotations and inversions associated to all the symmetries of the $n$-side
regular polygon; in other words, $D_n=\{I, R(2\pi/n), \ldots, R(2(n-1)\pi/n),
M_1, \ldots, M_n \}$, where $M_j$ is the mirror reflection on the $j$th
polygon symmetry axis.

It has been found~\cite{generalstates} that a set of $n$ orthogonal states
associated to the cyclic and dihedral groups can be defined, using an initial,
non-invariant state and its rotations plus reflections in the phase space.
These states are defined by the sum of the rotated and inverted states, where the weight of each element in the sum is
the character of the cyclic group representation.

The aim of this work is to define the Wigner functions and tomographic
probability representation of the superposition of Gaussian states associated
to the cyclic and dihedral groups defined in~\cite{generalstates}.

This paper is organized as follows.

In section~2, we review the approach of the tomographic representation of
quantum mechanics from the viewpoint of the quantizer--dequantizer formalism.
In section~3, we describe the symplectic-tomography representation for the
superposition of coherent states associated to the $C_3$~group as an example.
In section~4, we present the explicit expressions for the Wigner function and
the tomogram for general superposition of Gaussian states, which have the
cyclic and dihedral symmetries in the phase space; also here some general
properties are discussed. Finally, in section~4, we present a summary of our
work and concluding remarks.


\section{Quantizer--dequantizer formalism}
In order to describe the density operator by a function, which is a
probability distribution, we consider the general method of an invertible
mapping of operators acting in a Hilbert space onto functions called symbols
of the operators. For an operator $\hat A$, let the set of operators $\hat
U(x)$, called dequantizers, where $x=x_1,x_2,\ldots,x_n$, to provide the
function $f_A(x)$, due to the relationship
\begin{equation}\label{1}
f_A(x)=\mbox{Tr}\left(\hat A\hat U(x)\right).
\end{equation}
Let the other set of operators $\hat D(x)$, called quantizers, to provide the
inverse relationship
\begin{equation}\label{2}
\hat A=\int f_A(x)\hat D(x)\, dx.
\end{equation}
The existence of a pair of operators $\hat U(x)$ and $\hat D(x)$ means that
these operators satisfy the following condition for any operator $\hat A$:
\begin{equation}\label{3}
f_A(x')=\int f_A(x)\mbox{Tr}\left(\hat U(x')\hat D(x)\right)\,dx;
\end{equation}
a possibility to fulfill this condition is
\begin{equation}\label{4}
\mbox{Tr}\left(\hat U(x')\hat D(x)\right)=\delta(x-x').
\end{equation}
Although it is not unique. For example, in the case of symplectic tomographic
probability representation~\cite{ManciniPLA} of harmonic-oscillator states,
one can describe the system with three variables $x=(x_1,x_2,x_3)$, where
$x_1=X$, $x_2=\mu$, and $x_3=\nu$ are real numbers, and the quantizer and
dequantizer operators are defined as
\begin{eqnarray}
\hat U\left(X,\mu,\nu\right)=\delta \left(X\hat 1-\mu\hat q-\nu\hat
p\right),\label{5}\\
\hat D\left(X,\mu,\nu\right)=\frac{1}{2\pi}\,\exp\left[i\left(X\hat 1-\mu\hat
q-\nu\hat p\right)\right],\label{6}
\end{eqnarray}
with $\hat q$ and $\hat p$ being the position and momentum operators.

In this notation, the symbol (function associated to the density operator)
corresponding to the density operator is $w\left(X\mid\mu,\nu\right)$; it is
called symplectic tomogram and it can be written as
\begin{equation}
w\left(X\mid\mu,\nu\right)=\mbox{Tr}(\hat\rho\,\delta \left(X\hat 1-\mu\hat
q-\nu\hat p\right)).\label{8}
\end{equation}
The symplectic tomogram can be used to reconstruct the density operator in view of
the quantizer operator described above, namely,
\begin{equation}\label{10}
\hat\rho=\frac{1}{2\pi}\, \int w\left(X\mid\mu,\nu\right)\,\exp\left[i\left(X\hat
1-\mu\hat q-\nu\hat p\right)\right]\,dX\,d\mu\,d\nu.
\end{equation}
For the density operator $\hat\rho=\vert \psi\rangle\langle\psi\vert$ of the pure
state $\mid\psi\rangle$, the tomogram $w\left(X\mid\mu,\nu\right)$ is
expressed in terms of the wave function $\psi(y)$ as follows~\cite{MendesPLA}:
\begin{equation}\label{11}
w_\psi\left(X\mid\mu,\nu\right)=\frac{1}{2\pi|\nu|}\,\left|\int\psi(y)
\,\exp\left[i\left(\frac{\mu y^2}{2\nu}-\frac{Xy}{\nu}
\right)\right]\,dy\right|^2.
\end{equation}

Recalling the physical meaning of the tomogram, we state that, for a
normalized wave function $\psi(y)$, i.e., $\displaystyle{\int}
|\psi(y)|^2\,dy=1$, the tomogram is a nonnegative normalized probability
distribution of the variable $X$ depending on real parameters $\mu$ and $\nu$;
i.e., one has
\begin{equation}\label{12}
\int w\left(X\mid\mu,\nu\right)\,dX=1.
\end{equation}
The real parameters $\mu=s\cos\theta$ and $\nu=s^{-1}\sin\theta$, determine the
reference-frame of transformed axes in the phase space $\hat q'=\mu\hat q+\nu\hat
p$ and $\hat p'=-\nu \hat q+\mu \hat p$, such that $\left[\hat q',\hat
p'\right]=\left[\hat q,\hat p\right]=i\hat 1$; $\hbar=1$, which implies that the matrix
$\left(\begin{array}{cc} \mu & \nu \\
 -\nu & \mu \end{array}\right)$ is symplectic.

A classical analog of the symplectic tomogram is determined by the probability
distribution $f(q,p)$ of a particle in the phase space, in view of the Radon
transform, namely,
\begin{eqnarray}
w_{\rm cl}\left(X\mid\mu,\nu\right)=\int f(q,p)\,\delta \left(X-\mu q-\nu
p\right)\,dq\,dp,\label{13}\\
f(q,p)=\frac{1}{4\pi^2}\,\int w_{\rm cl}\left(X\mid\mu,\nu\right)
\,\exp\left[i\left( X-\mu q -\nu p \right)\right]\,dX\,d\mu\,d\nu.\label{14}
\end{eqnarray}
The classical particle tomogram $w_{\rm cl}\left(X\mid\mu,\nu\right)$
determines the probability distribution $f(q,p)$ and, being an analog of the
quantum particle tomogram determined by the Wigner function, it reads
\begin{eqnarray}
w\left(X\mid\mu,\nu\right)=\int W(q,p)\,\delta \left(X-\mu
q-\nu p\right)\frac{dq\,dp}{2\pi}\,,\label{15}\\
W(q,p)=\frac{1}{2\pi}\int w\left(X\mid\mu,\nu\right) \,\exp\left[i\left( X-\mu
q -\nu p \right)\right]\,dX\,d\mu\,d\nu.\label{16}
\end{eqnarray}
Here, for the particle pure state $W_\psi$, the Wigner function is
\begin{equation}
W_\psi(q,p)=\frac{1}{2\pi}\int
 \psi\left(q+u/2\right) \psi^*\left(q-u/2\right)e^{-ipu} du,
\label{17}
\end{equation}
and it satisfies the normalization condition $\displaystyle{\int}
W_\psi(q,p)\,\dfrac{dq\,dp}{2\pi}=1$, for the normalized pure state.

In order to exemplify the standard procedure, we present an explicit
definition of tomographic probability representation for the superposition of
coherent states associated to the triangle rotation symmetries in the phase
space.


\section{Crystallized cat states on an example of the $C_3$ symmetry group }

In~\cite{DodPhysica1974}, the even and odd coherent states were introduced,
using the symmetry group $C_2$,  which is a set of the abelian group
transformations applied to the Glauber coherent states $\mid\alpha\rangle$ of
the harmonic oscillator; they provide two states (even and odd)
\begin{eqnarray}
\mid\alpha_+\rangle = N_+\left(\mid\alpha\rangle + \mid -
\alpha\rangle\right),\qquad
N_+=\left[2\left(1+e^{-2|\alpha|^2}\right)\right]^{-1/2},\label{18}\\
\mid\alpha_-\rangle = N_-\left(\mid\alpha\rangle - \mid -
\alpha\rangle\right),\qquad
N_-=\left[2\left(1-e^{-2|\alpha|^2}\right)\right]^{-1/2},\label{19}
\end{eqnarray}
satisfying the equation
\begin{equation}
\hat a^2 \mid\alpha_\pm\rangle=\alpha^2\mid\alpha_\pm\rangle,\qquad \hat
a=\frac{\hat q+ i\hat p}{\sqrt2}\,.\label{20}
\end{equation}

The states under consideration can be generalized using the abelian symmetry
group $C_3$ with three rotation elements $1$, $e^{2\pi i/3}$, and $e^{4\pi
i/3}$ acting on the coherent states. It is the simplest generalization of the
$C_2$ symmetry group, used for the definition of even and odd coherent
states~\cite{DodPhysica1974} and being the rotation group $C_3$. In view of
this symmetry group, we define the superposition of coherent states
$\mid\alpha\rangle$ of the form
\begin{equation}
\mid\psi\rangle=N_3\,\big(\mid\alpha\rangle + \mid\alpha e^{2\pi i/3}\,
\rangle + \mid\alpha e^{4\pi i/3}\,\rangle\,\big)\,,\label{23}
\end{equation}
where the identity irreducible representation of this group is used to obtain
the coefficients in the sum of coherent states. The state $ \vert \psi\rangle$
is a sum of three Gaussian states, i.e., the states with normalized wave
functions $\langle\psi_i\mid\psi_j\rangle=1$, given in the position
representation by the expressions
\begin{equation}
\langle x\mid\psi_j\rangle=\psi_j(x)=\exp\left(A_jx^2+B_jx+C_j\right); \quad
j=1,2,3,\label{24}
\end{equation}
and the superposition state vector reads
\begin{equation}
\mid\psi\rangle=\sum_{j=1}^3D_j\mid\psi_j\rangle . \label{25}
\end{equation}
The normalization condition $\langle\psi\mid\psi\rangle=1$ means that, for
normalized states $\psi_j(x)$ giving the relations
\begin{equation}
C_j+C_j^*=\frac12\,\ln\left(\frac{A_j+A_j^*}{\pi}\right)-\frac{(B_j+B_j^*)^2}{4
(A_j+A_j^*)}\,,\label{26}
\end{equation}
the normalization of state~(\ref{25}) provides the connection for coefficients
following from the equality
\begin{equation}
\langle\psi\mid\psi\rangle=\sum_{j,k=1}^3D_jD_k^*\langle\psi_j\mid\psi_k\rangle=1.
\label{27}
\end{equation}
For states~(\ref{24}), the coefficients are
\begin{equation}
A_j=\frac12,\quad B_j = \sqrt2\alpha e^{(2\pi i/3)j},\quad C_j=
-\frac{|\alpha|^2}{2}-\frac{[\alpha e^{(2\pi i/3)j}]^2}{2} \,\quad
D_j=N_3;\quad j=1,2,3.\label{28}
\end{equation}
Our aim is to obtain the tomographic probability distribution
$w_\psi(X\mid\mu,\nu)$ for the state with the wave function $\psi(x)$
corresponding to the state vector~(\ref{23}).

Adopting the generic relation from~\cite{MendesPLA}, we write the tomogram in
terms of Gaussian integrals as follows:
\begin{equation}
w_\psi(X\mid\mu,\nu)=\frac{1}{2\pi|\nu|}\left|\int
dy\sum_{j=1}^3D_j\exp\left(-A_jy^2+B_jy+C_j\right) \exp\left(\frac{i\mu}{2
\nu}y^2-\frac{iX}{\nu}y\right)\right|^2,\label{29}
\end{equation}
where $A_j$, $B_j$, $C_j$, and $D_j$ are given by (\ref{28}). Then the
tomogram for the state~(\ref{23}) can be written as
\begin{eqnarray}
w_\psi(X\mid\mu,\nu)&=&\frac{1}{2\pi|\nu|}\left|N_1e^{C_1}
\sqrt{\frac{\pi}{A_1-i\mu/2
\nu}}~\exp\,\frac{(B_1-iX/2\nu)^2}{4(A_1-i\mu/2\nu)}\right.\nonumber\\
&&~\quad +\left.N_2e^{C_2}\sqrt{\frac{\pi}{A_2-i\mu/2
\nu}}~\exp\,\frac{(B_2-iX/2\nu)^2}{4(A_2-i\mu/2\nu)}\right.\nonumber\\
&&~\quad +\left.N_3e^{C_3}\sqrt{\frac{\pi}{A_3-i\mu/2
\nu}}~\exp\,\frac{(B_3-iX/2\nu)^2}{4(A_3-i\mu/2\nu)}\right|^2. \label{29-1}
\end{eqnarray}
Here, the coefficients $N_j$,$A_j$,$B_j$,$C_j$; $j=1,2,3$ are given as
functions of complex parameters $\alpha$ describing Schr\"odinger cat states.
In the case of an arbitrary superposition of $N$ Gaussian states
$\mid\psi\rangle=\sum_{j=1}^NN_j\mid\psi_j\rangle$, with the normalized state
vector $\mid\psi_j\rangle$, the general formula for tomogram is a sum of
normalized Gaussian terms,
\begin{eqnarray}
w_\psi(X\mid\mu,\nu)&=&\frac{1}{2\pi|\nu|}\left|\sum_{j=1}^N N_je^{C_j}
\sqrt{\frac{\pi}{A_j-i\mu/2
\nu}}~\exp\,\frac{(B_j-iX/2\nu)^2}{4(A_j-i\mu/2\nu)} \right|^2. \label{30}
\end{eqnarray}
These tomograms describe also states, which are the states of crystallized
Schr\"odinger cat states corresponding to an arbitrary symmetry group.


\section{General superpositions of Gaussian states associated to the cyclic and dihedral groups}

The superposition of coherent states studied above is a particular case of a
symmetric superposition of states. One can define more general cyclic or
dihedral states by considering an initial state $\vert \phi \rangle$, which we
assume to be non-invariant under rotations in the phase space (to define
cyclic states) or non-invariant under rotations and inversions in the phase
space (to define dihedral states)~\cite{generalstates}.

Given this non-invariant state $\vert \phi \rangle$, which is required to be non-invariant under
the rotations in the phase space for now, we can define $n$ states associated to the
cyclic group of $n$ degree as follows~\cite{generalstates}:
\begin{equation}
\vert \psi_r^{(\lambda)} (\phi) \rangle = N_\lambda \sum_{j=1}^n
\chi^{(\lambda)}_r \hat{R}(\theta_r) \vert \phi \rangle,
\end{equation}
with $\lambda$ being one of the $n$ irreducible representations of the cyclic
group. Here, the rotation operator is denoted by $\hat{R}(\theta_r)=e^{-i
\theta_r \hat{n}}$ and the character of the group is $\chi_r^{(\lambda)}=e^{{2
\pi i(\lambda-1)(r-1)}/{n}}$, with the rotation angle $\theta_j= 2 \pi(j-1)/n$
and the normalization constant
\begin{equation}
N_\lambda^{-2}=\sum_{r,s=1}^n \chi_r^{(\lambda)} \chi_s^{(\lambda)*} \langle
\phi\vert \hat{R}^\dagger(\theta_s) \hat{R}(\theta_r) \vert \phi \rangle.
\end{equation}
This state has an associated Wigner function, which can be obtained by the
following integral:
\[
W_\lambda(x,p)=\frac{1}{2 \pi \hbar}\int d \xi \, e^{-({i}/{\hbar})\, p \xi}\,
\psi^*_\lambda \left(x-\frac{1}{2}\xi\right) \psi_\lambda
\left(x+\frac{1}{2}\xi\right).
\]
In the case of the cyclic states, $W_\lambda(x,p)$ takes the form
\begin{equation}
W_\lambda(x,p)=\frac{N_\lambda^2}{2\pi \hbar} \sum_{r,s=1}^n
\int_{-\infty}^\infty d\xi \, e^{-({i}/{\hbar})\, p \xi} \chi_r^{(\lambda)*}
\chi_s^{(\lambda)} \phi_r^* \left(x-\frac{1}{2}\xi\right) \phi_s
\left(x+\frac{1}{2}\xi\right), \label{wigner_cyc}
\end{equation}
with $\phi_j(x)=\langle x \vert \hat{R}(\theta_j) \vert \phi \rangle$.

For the cyclic Gaussian states \cite{generalstates} associated to the initial
non-invariant state
\begin{equation}
\phi(x)=\left( \frac{a+a^*}{\pi}\frac{1+2a}{1+2a^*}\right)^{1/4}
e^{-\frac{b^2+ \vert b \vert^2}{4(a+a^*)}}~e^{-a x^2+ b x}, \label{gauss}
\end{equation}
one can demonstrate that, in the position representation, its rotation in the
phase space $\langle x \vert \hat{R}(\theta_j) \vert \phi \rangle$ is given by
the following expression:
\begin{equation}
\phi_j(x)=\langle x \vert \hat{R}(\theta_j)\vert \phi \rangle= \left(
\frac{a_j+a^*_j}{\pi}\frac{1+2a_j}{1+2a^*_j}\right)^{1/4} e^{-\frac{b^2_j+
\vert b_j \vert^2}{4(a_j+a^*_j)}}~e^{-a_j x^2+ b_j x},
\end{equation}
where the parameters
\[
a_j = \frac{2 i a \cos \theta_j-\sin \theta_j}{2(i \cos \theta_j -2 a \sin
\theta_j)}, \quad b_j =\frac{b}{\cos \theta_j +2i a \sin \theta_j}
\]
are taken into account. The normalization constant, in this case, is
\begin{equation}
N_\lambda^{-2}= \sum_{r,s=1}^n c_r c_s^* ~\sqrt{\frac{\pi}{a_r+a_s^*}}~
e^{\frac{(b_r+b_s^*)^2}{4(a_r+a_s^*)}}, \quad c_j= \chi_j^{(\lambda)}
 \left( \frac{a_j+a^*_j}{\pi}\frac{1+2a_j}{1+2a^*_j}\right)^{1/4}
 e^{-\frac{b^2_j+ \vert b_j \vert^2}{4(a_j+a^*_j)}}.
\end{equation}

The integral part of the Wigner function of Eq.~(\ref{wigner_cyc}) can be
obtained using the integral, defined as
$W_{r,s}(x,p)=\displaystyle{\int}_{-\infty}^\infty d\xi \, e^{-({i}/{\hbar})\,
p \xi} \chi_r^{(\lambda)*} \chi_s^{(\lambda)} \phi_r^* \left(x-\xi/2\right)
\phi_s \left(x+\xi/2\right)$, which reads
\begin{equation}
W_{r,s}(x,p)=2~\sqrt{\frac{\pi}{a_r+a_s^*}}~c_r^*\,c_s~e^{-\frac{4 a_r^*
a_s}{a_r^*+a_s} x^2-\frac{p^2}{(a_r^*+a_s)\hbar^2}+\frac{2 i p x
(a_s-a_r^*)}{\hbar (a_r^*+a_s)}+ \frac{2 (a_r^* b_s+a_s b_r^*)}{a_r^*+a_s} x+
\frac{i p (b_r^*-b_s)}{\hbar (a_r^*+a_s)} +\frac{(b_r^*-b_s)^2}{4
(a_r^*+a_s)}}. \label{int1}
\end{equation}
The results obtained provide the possibility to obtain the cyclic Wigner
function as follows:
\begin{equation}
W_\lambda(x,p)= \frac{N_\lambda^2}{2 \pi \hbar}\, \sum_{r,s=1}^n W_{r,s}(x,p).
\end{equation}

\begin{figure}
\centering
\includegraphics[scale=0.4]{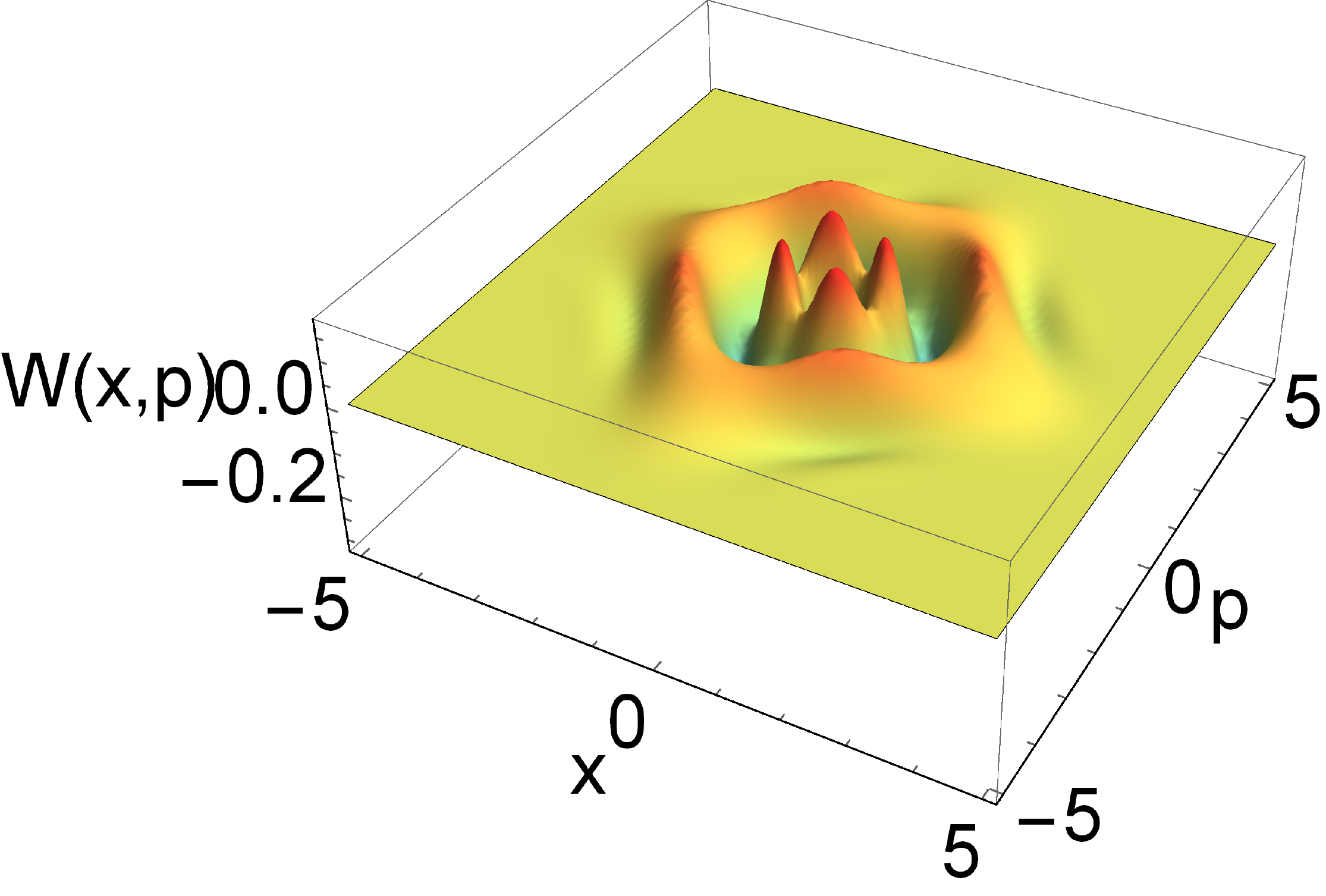} \
\includegraphics[scale=0.25]{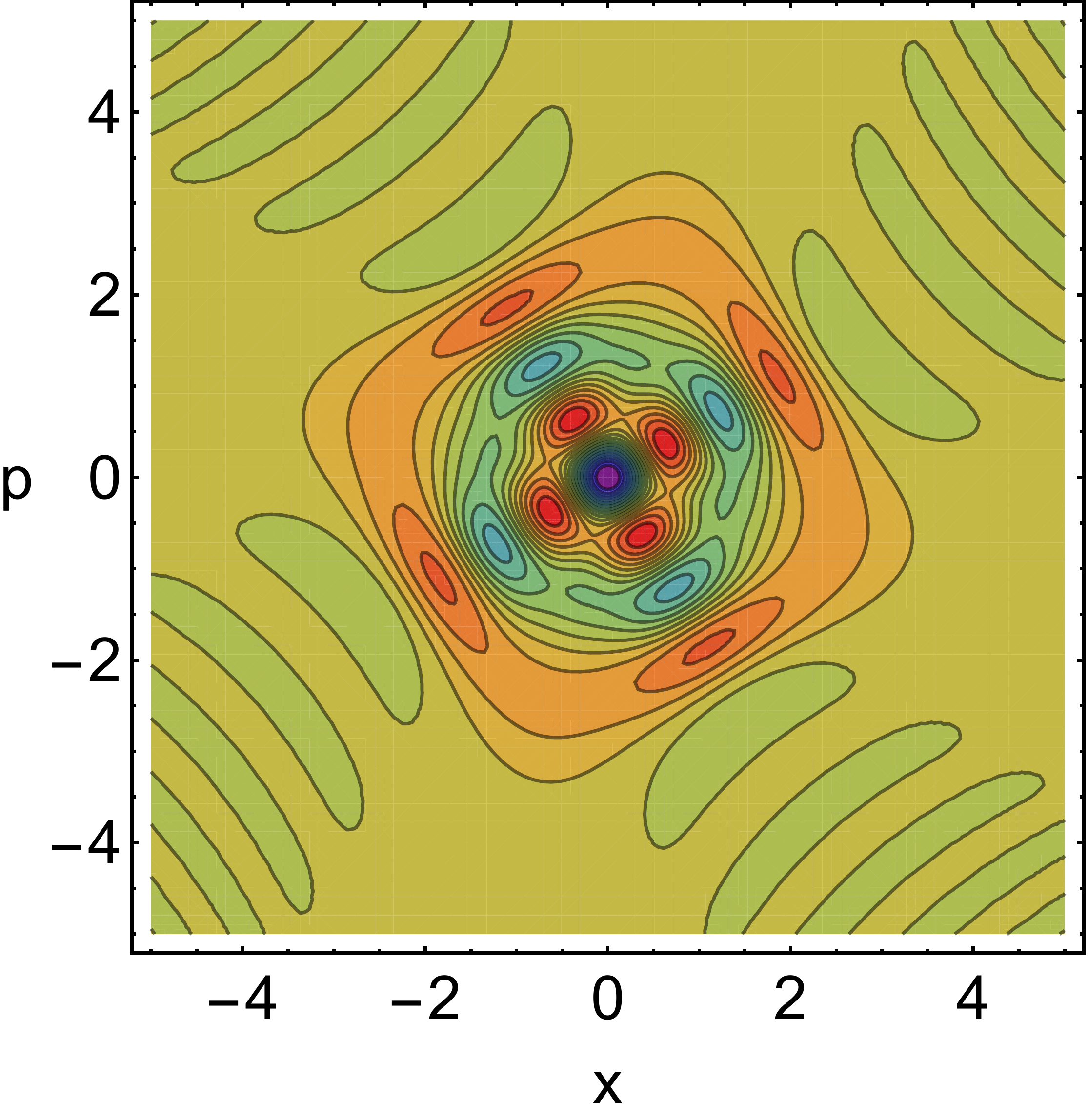}
\caption{Wigner function for the cyclic state of fourth order with irreducible
representation $\lambda=1$~(left)  and its contour plot~(right). Here, the
parameters $a=1$ and $b=\sqrt{2}(1+i)$ are taken.\label{fourth}}
\end{figure}

\begin{figure}
\centering
\includegraphics[scale=0.4]{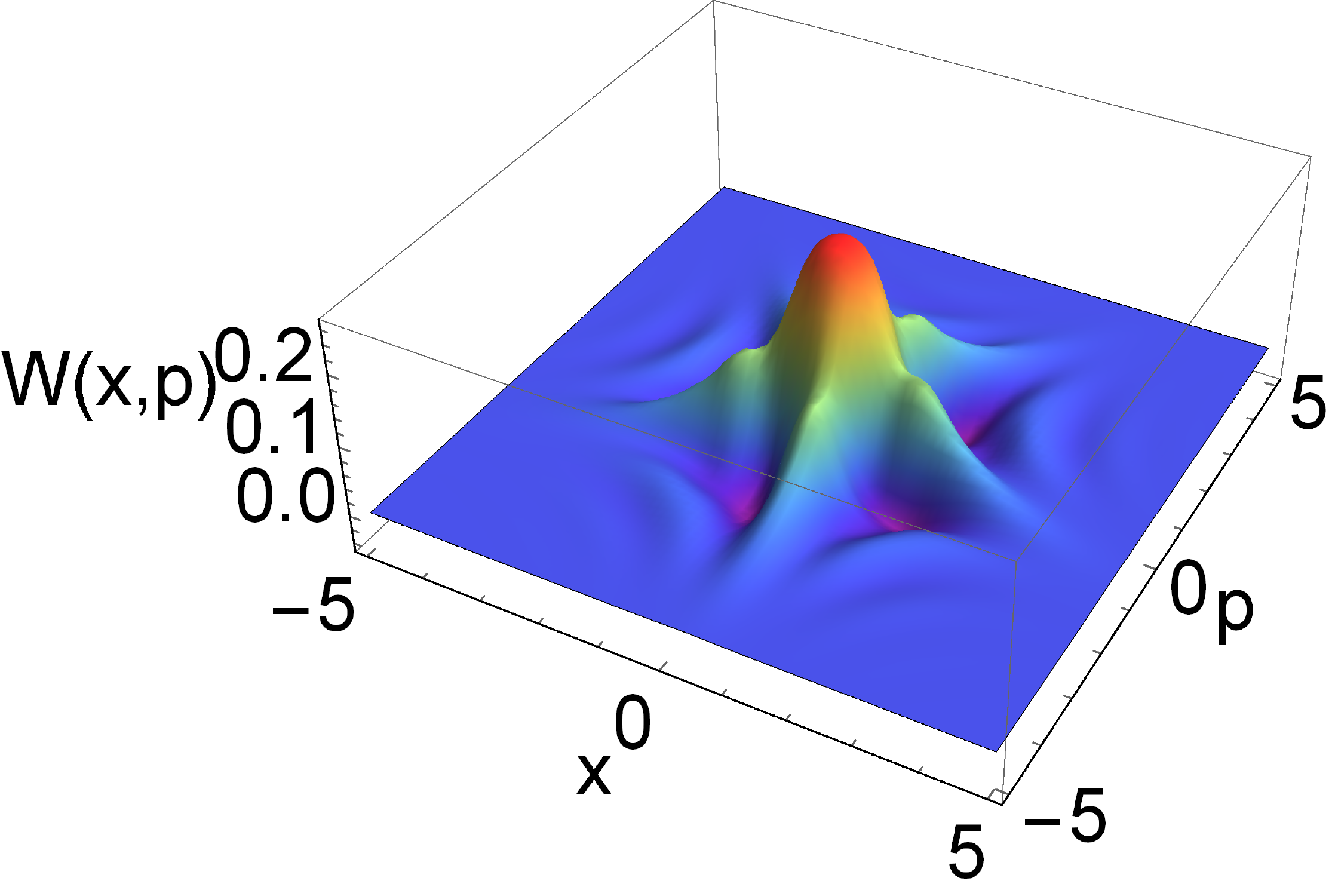} \
\includegraphics[scale=0.25]{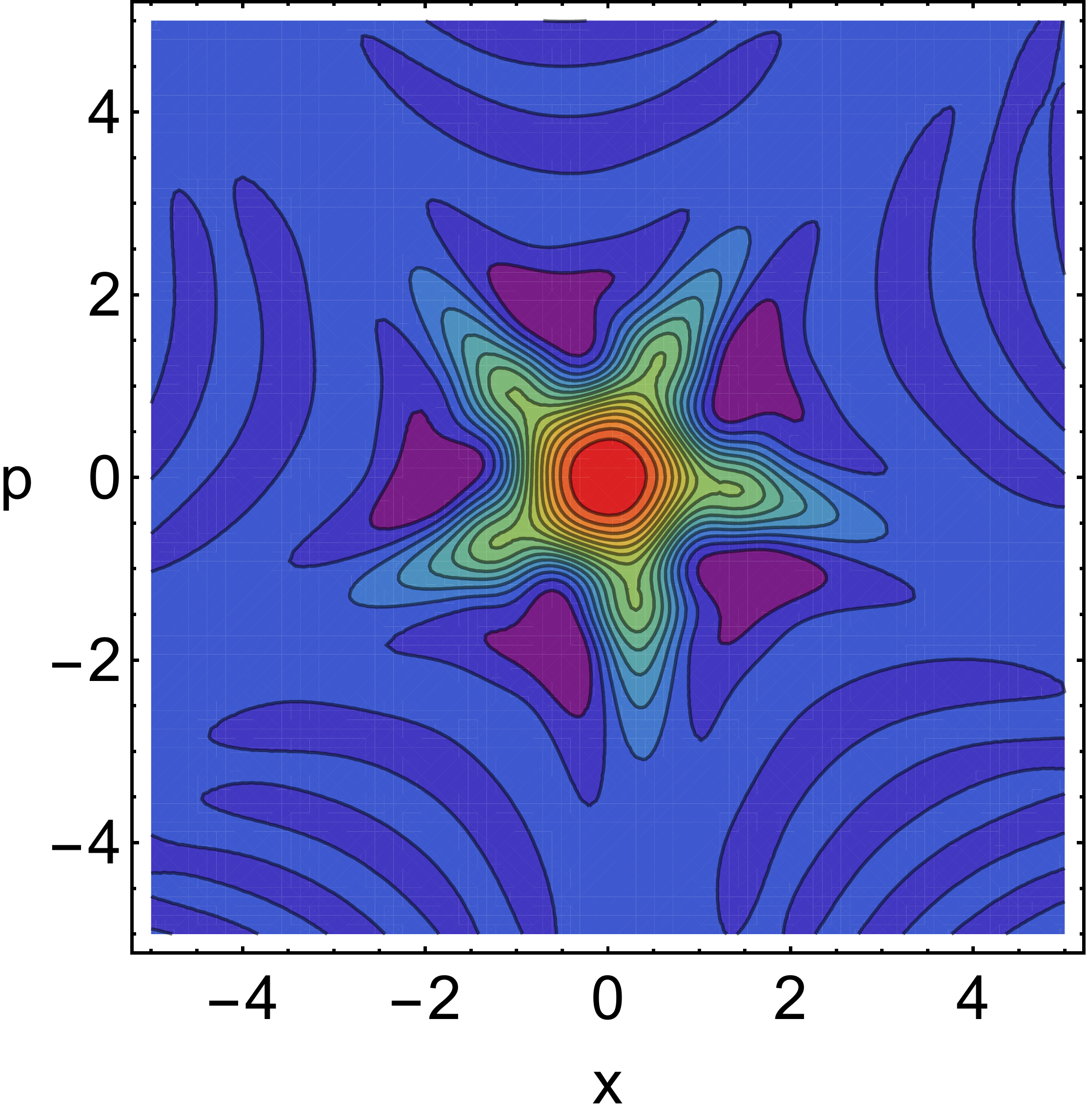}
\caption{Wigner function for the cyclic state of fifth order with irreducible
representation $\lambda=1$~(left) and its contour plot~(right). Here, the
parameters $a=2$ and $b=\sqrt{2}(1-i)$ are taken.\label{fifth}}
\end{figure}

In Fig.~\ref{fourth}, the Wigner function for the cyclic state associated to
the cyclic group $C_4$ and irreducible representation $\lambda=4$ is shown. In
this figure. one can confirm the rotation symmetry in the phase space for
rotations multiple of $\pi/2$: it is also seen that the inversion symmetry is
not present as that symmetry is not present in the cyclic group. Analogously,
the pentagon rotation symmetry can be seen in the Wigner function presented in
Fig.~\ref{fifth}.

In addition to the cyclic states, we present the Wigner function for the
dihedral states of degree $n$, which have the same symmetries as the dihedral
group $D_n$ in the phase space. In other words, they are invariant under the
rotations of the cyclic group and also under certain inversions, just like a
regular polygon of $n$ sides. The dihedral states are defined
as~\cite{generalstates}
\begin{equation}
\vert \gamma_r^{(\lambda)} (\phi) \rangle = N_\lambda \sum_{r=1}^n (\chi_r^{(\lambda)}
\hat{U}_r \vert \phi \rangle+\chi_r^{(\lambda)*}\,  \hat{U}_r^* \vert \phi^* \rangle),
\end{equation}
with the dihedral group operator $\hat{U}_r=\hat{C} \hat{R}(\theta_r)$ being
composed of a rotation $\hat{R}$ and the complex conjugation $\hat{C}$. The
characters $\chi_r^{(\lambda)}=e^{\frac{2 \pi i(\lambda-1)(r-1)}{n}}$ are the
ones for the cyclic subgroup. Then the Wigner function associated to these
states is defined as
\begin{equation}
W(x,p)=\frac{N_\lambda^2}{2 \pi \hbar} \sum_{r,s=1}^n \int_{-\infty}^\infty d
\xi \, e^{-i p \xi / \hbar}\left\langle x- \frac{1}{2}\xi \right\vert
(\chi_r^{(\lambda)} \hat{U}_r \vert \phi \rangle + \chi_r^{(\lambda)*} \,
\hat{U}_r^\dagger \vert \phi^* \rangle)(\chi_s^{(\lambda)*}\langle \phi \vert
\hat{U}_s^\dagger+ \chi_s^{(\lambda)} \langle \phi^* \vert \hat{U}_s)
\left\vert x+\frac{1}{2}\xi \right\rangle.
\end{equation}
In the case of Gaussian dihedral states defined by the initial non-invariant
state in Eq.~(\ref{gauss}), the Wigner function can be calculated with the
help of the integral of Eq.~(\ref{int1}) along with the following integrals:
\begin{eqnarray}
W_{r,s}^\prime (x,p)=\int_{-\infty}^\infty d\xi \, e^{-i p \xi /\hbar} \,
\phi_r^*\left(x-\frac{1}{2}\xi \right) \phi_s^* \left(x+\frac{1}{2}\xi \right), \\
W_{r,s}^{\prime \prime} (x,p)=\int_{-\infty}^\infty d\xi \, e^{-i p \xi /\hbar} \,
\phi_r\left(x-\frac{1}{2}\xi \right) \phi_s^* \left(x+\frac{1}{2}\xi \right), \\
W_{r,s}^{\prime \prime \prime}(x,p)=\int_{-\infty}^\infty d\xi \, e^{-i p \xi
/\hbar}\, \phi_r\left(x-\frac{1}{2}\xi \right) \phi_s \left(x+\frac{1}{2}\xi
\right).
\end{eqnarray}
In order to calculate these integrals, we point out that they are analogous to
Eq.~(\ref{int1}) with some replacements. For example, $W_{r,s}^\prime (x,p)$
is equal to the expression of $W_{r,s}(x,p)$ with the replacements:
$a_s\rightarrow a_s^*$, $a_s^* \rightarrow a_s$, $b_s\rightarrow b_s^*$, and
$b_s^* \rightarrow b_s$. In view of this type of identification, the Wigner
function for the dihedral states can be written as
\begin{equation}
W(x,p)= \frac{N_\lambda^2}{2 \pi \hbar} \sum_{r,s=1}^n \left(W_{r,s}(x,p)+
W_{r,s}^\prime(x,p)+W_{r,s}^{\prime \prime}(x,p)+W_{r,s}^{\prime \prime
\prime}(x,p)\right).
\end{equation}

\begin{figure}
\centering
\includegraphics[scale=0.4]{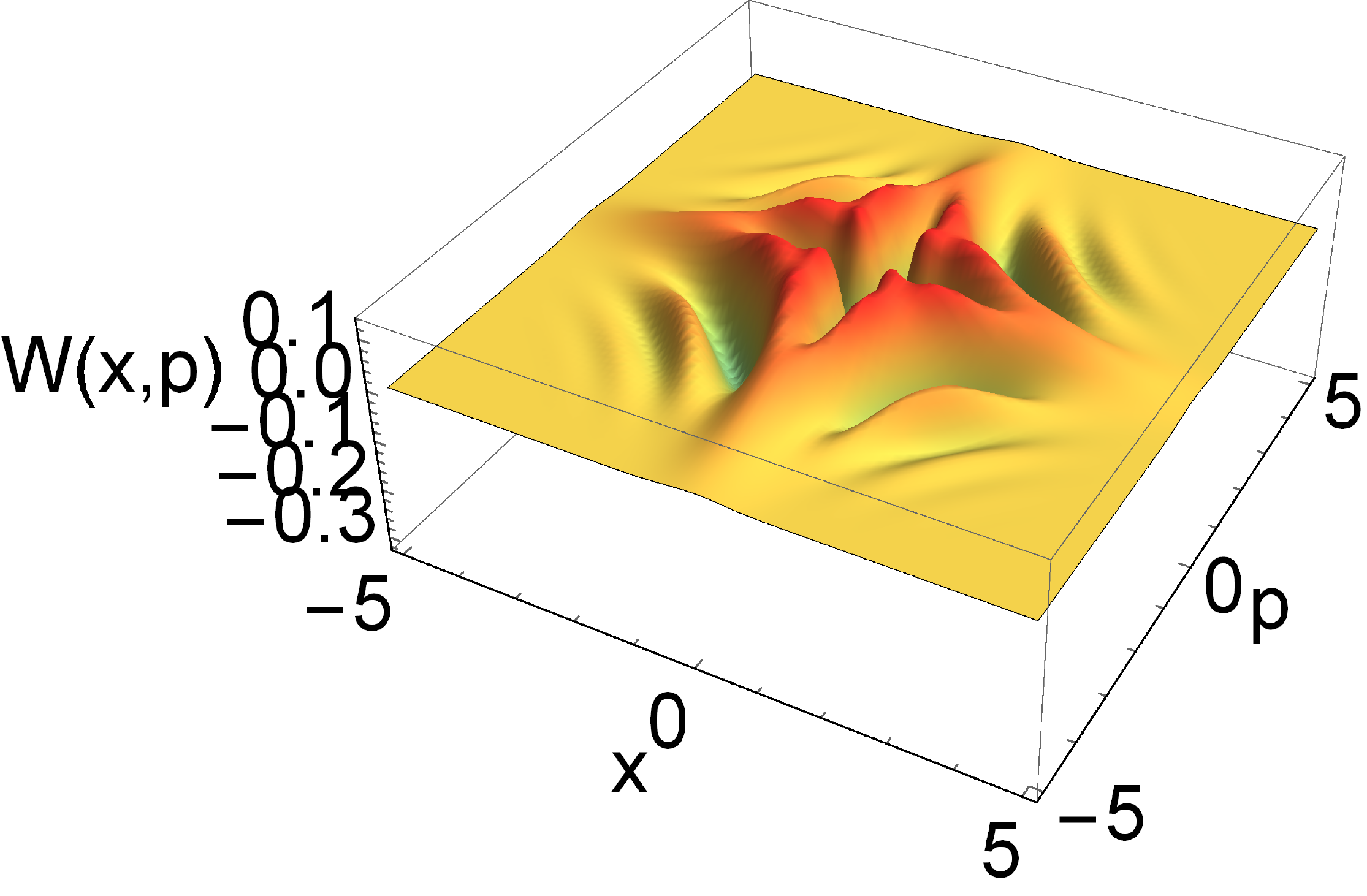} \
\includegraphics[scale=0.25]{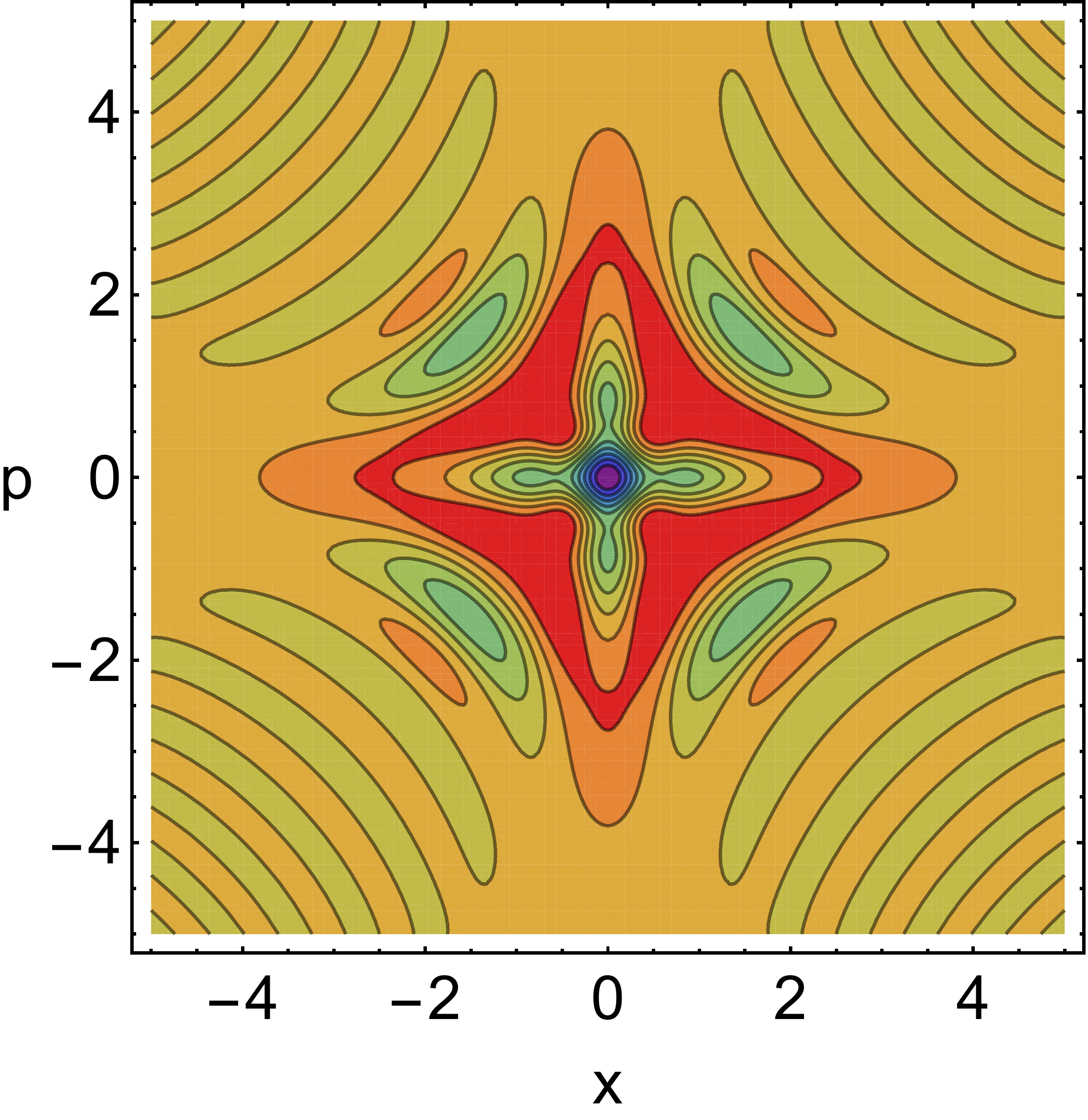}
\caption{Wigner function for the dihedral state of fourth order with
irreducible representation $\lambda=2$~(left) and its contour plot~(right). The parameters $a=1$ and $b=\sqrt{2}(1+i)$ were taken.\label{fourthd}}
\end{figure}

\begin{figure}
\centering
\includegraphics[scale=0.4]{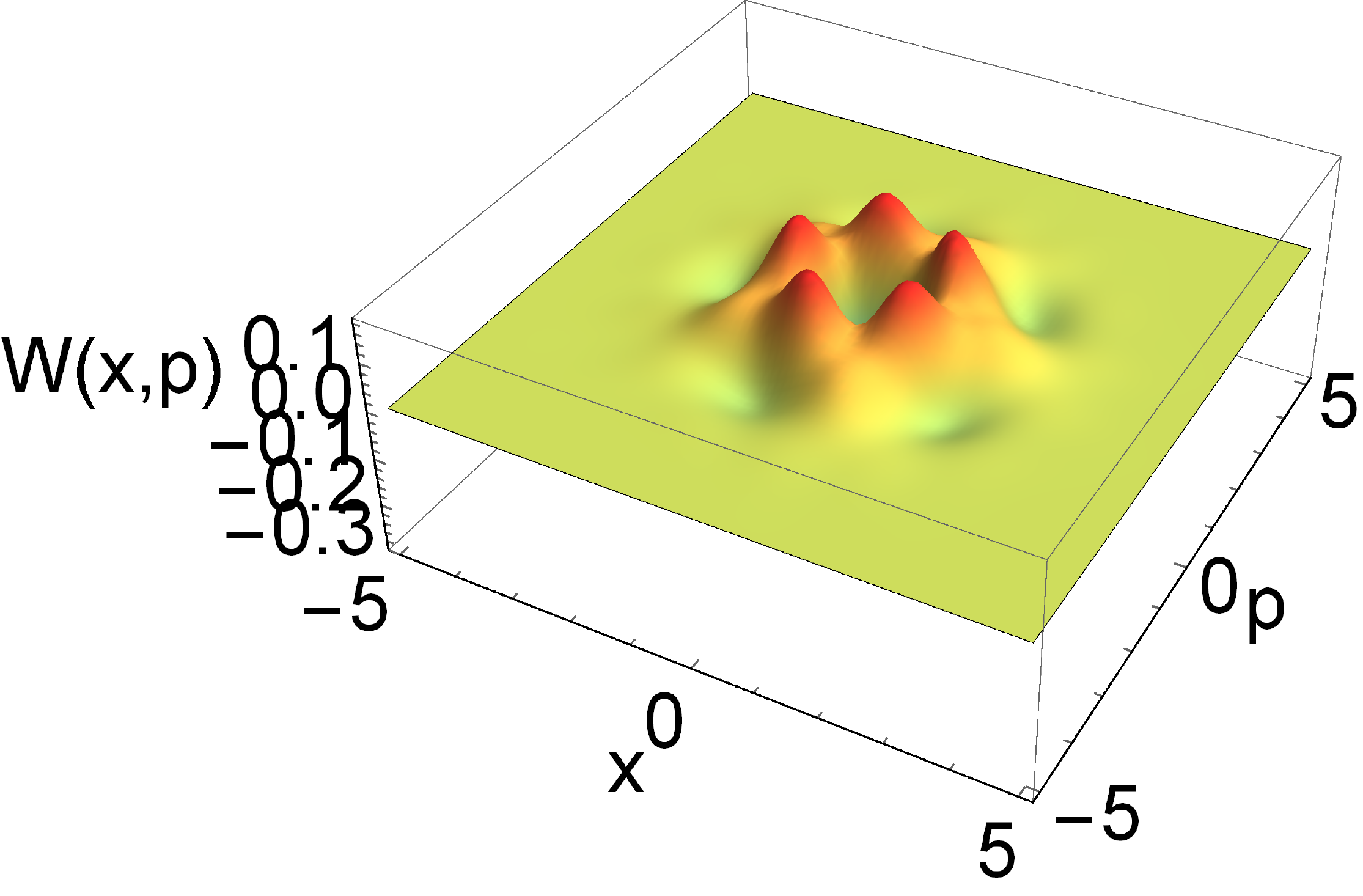} \
\includegraphics[scale=0.25]{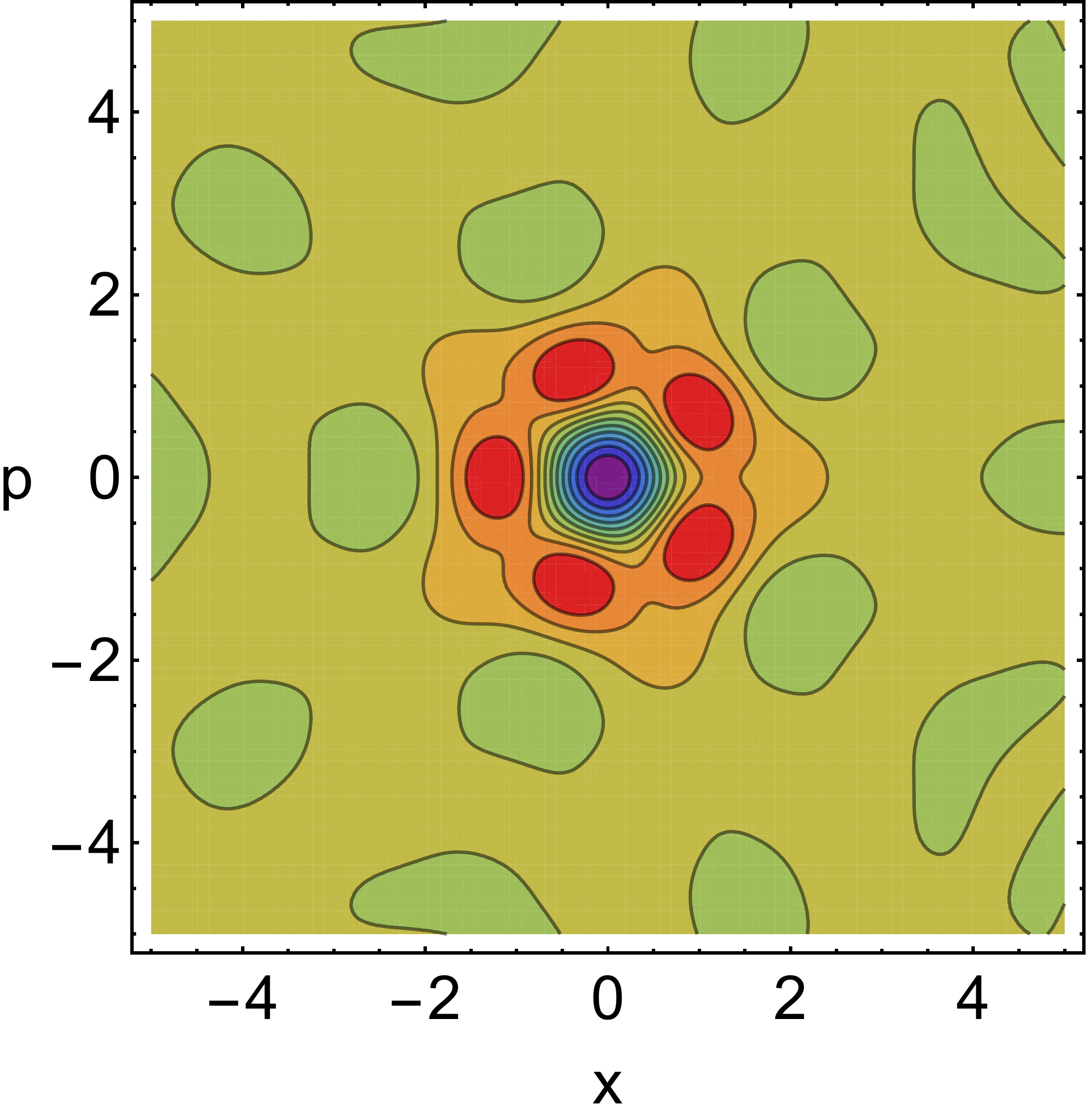}
\caption{Wigner function for the dihedral state of fifth order with
irreducible representation $\lambda=2$~(left) and its contour plot~(right).
Here, the parameters $a=2$ and $b=\sqrt{2}(1-i)$ were taken.\label{fifthd}}
\end{figure}

In Fig.~\ref{fourthd}, the Wigner function for the dihedral state of fourth
order ($D_4$) is shown. One can see that the obtained quasiprobability
distribution has both the rotational and inversion symmetries opposed to the
ones seen in Fig.~\ref{fourth} for the cyclic group of the same degree. In a
similar way, the Wigner function in Fig.~\ref{fifthd} has the pentagon
symmetries of $D_5$, both rotational and inversion instead of only the
rotational ones for the cyclic case seen in Fig.~\ref{fifth}.

\section{Tomographic representation for cyclic and dihedral states}
As previously stated, the tomogram of a quantum system is the probability
distribution for the position $X$  in a rotated and rescaled coordinate
system. In general, this tomogram can be calculated from the Wigner function
and, in the case of the cyclic states, the tomographic representation of the
states takes the form
\begin{equation}
w_\lambda(X,\theta)=\int_{-\infty}^\infty W_\lambda\,\big(X(x,p),P(x,p)\big)~
dP =\frac{N_\lambda^2}{2 \pi \hbar} \sum_{r,s=1}^n \int_{-\infty}^\infty
W_{r,s}\big(X(x,p),P(x,p)\big)~dP ,
\end{equation}
where $X(x,p)=x s \cos \theta+p s^{-1} \sin \theta$ and $P(x,p)=p s^{-1} \cos
\theta-x s \sin \theta$.  This integral provides the following result:
\begin{eqnarray}
w_\lambda(X,\theta)=\frac{N_\lambda^2}{2 \pi \hbar} \sum_{r,r'=1}^n w_{r,r'}(X, \theta),
\end{eqnarray}
with
\begin{eqnarray}
w_{r,r'}(X,\theta)=   \frac{2 \pi c_r^* c_{r'} \vert s \vert }{\sqrt{\left(s
^2 \cos \theta+2 i a_{r'} \sin \theta\right) \left(s ^2 \cos \theta-2 i \sin
\theta a_r^*\right)}}~ e^{-\frac{s ^2 X^2 \left(a_r^*+a_{r'}\right)}{\left(s
^2 \cos \theta+2 i a_{r'}
\sin \theta\right) \left(s ^2 \cos \theta-2 i \sin \theta a_r^*\right)}} \nonumber \\
\times e^{X \left(\frac{s  b_r^*}{s ^2 \cos \theta-2 i \sin \theta a_r^*}
+\frac{s b_{r'}}{s ^2 \cos \theta+2 i a_{r'} \sin \theta}\right) +\frac{1}{2}
\sin \theta \left(\frac{b_r^{*2}}{2 \sin \theta a_r^* +i s ^2 \cos
\theta}+\frac{b_{r'}^2}{2 a_{r'} \sin \theta-i s ^2 \cos \theta}\right)}\,.
\label{tomrs}
\end{eqnarray}

\begin{figure}
\centering
\includegraphics[scale=0.4]{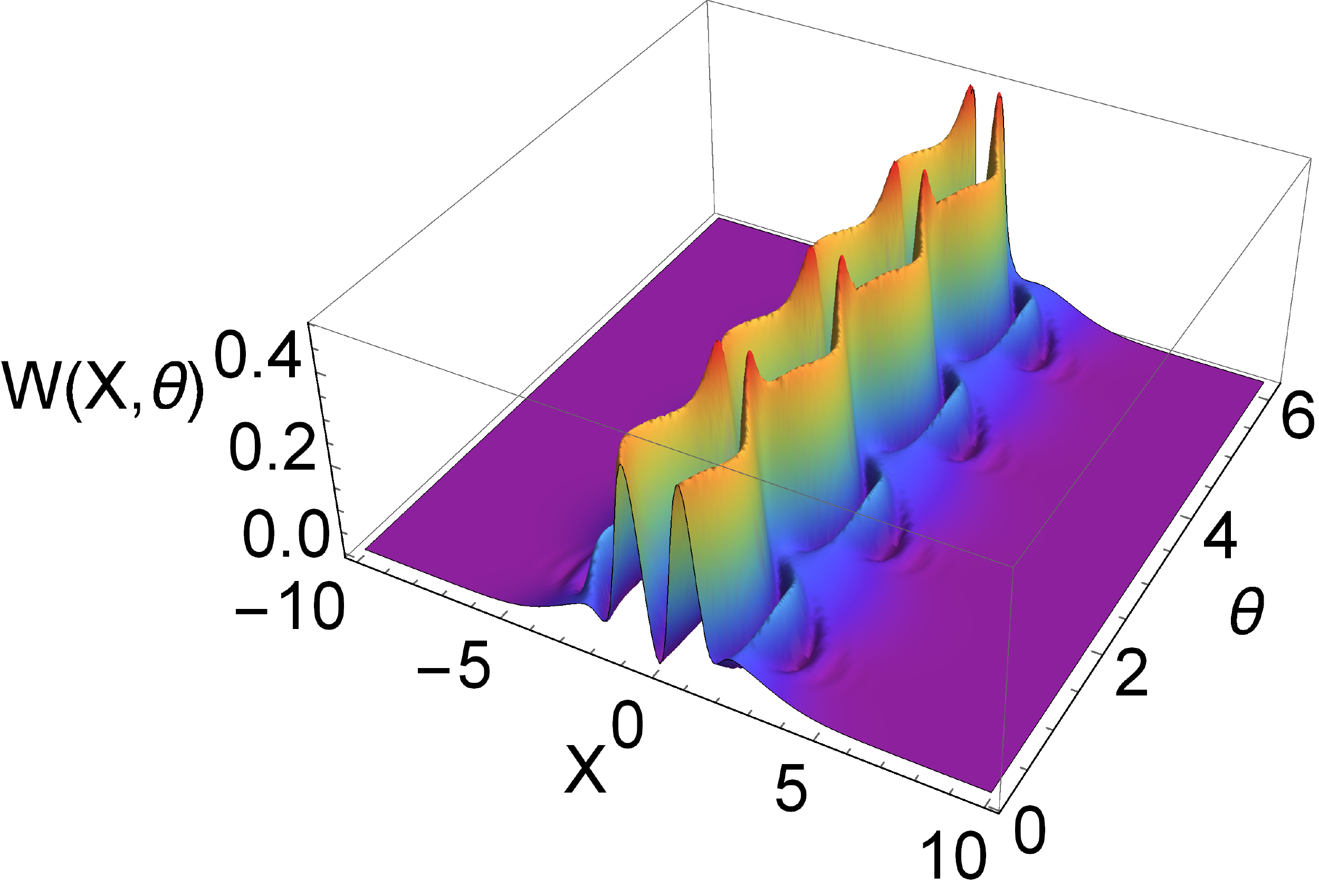} \
\includegraphics[scale=0.25]{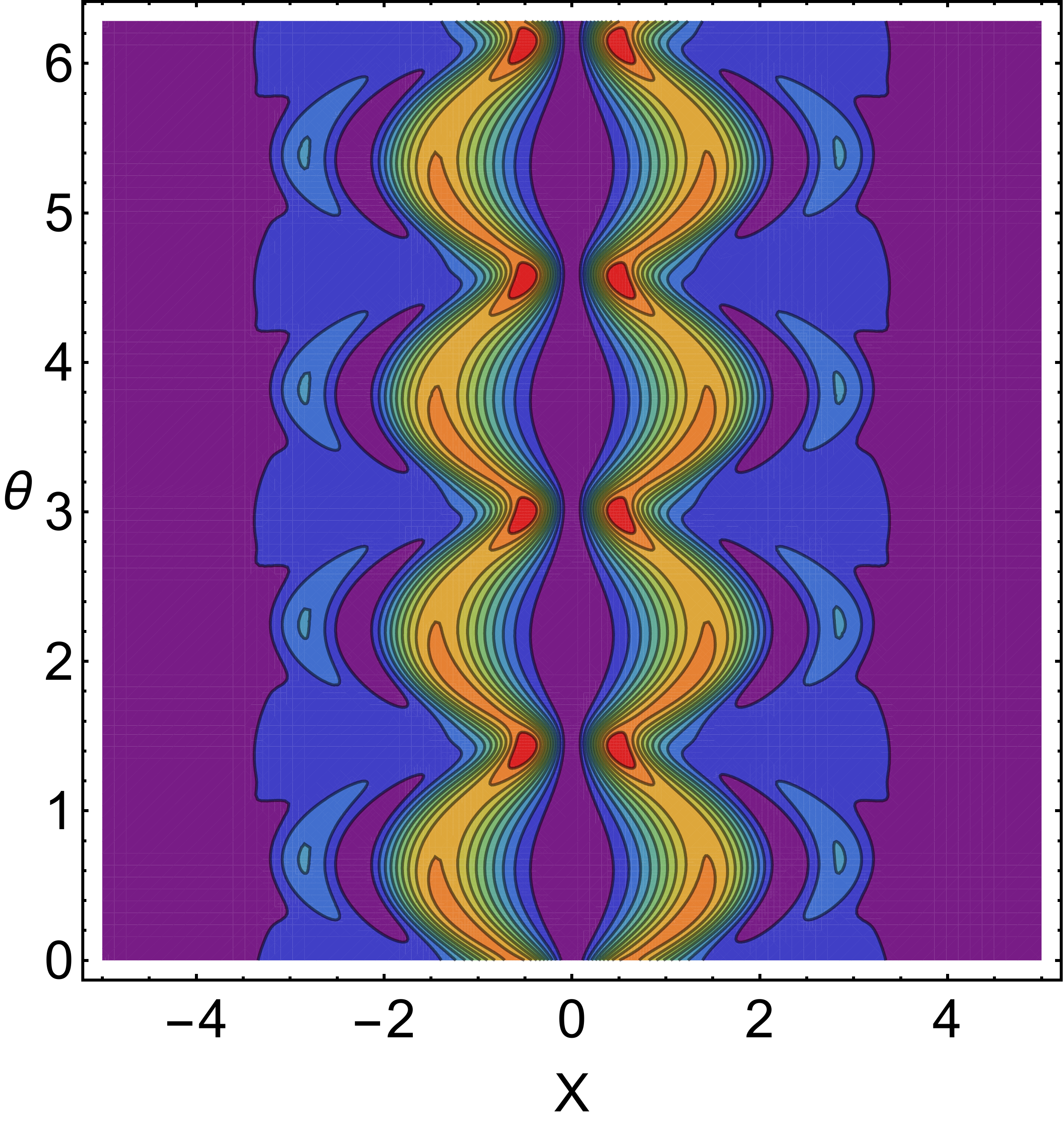}
\caption{Tomographic representation for the cyclic state of fourth order $C_4$
associated to the irreducible representation $\lambda=2$~(left) and its
contour plot~(right). The parameters $a=2$, $b=2(1+i)$, and $s=1$ were
considered.\label{cyc4}}
\end{figure}

\begin{figure}
\centering
\includegraphics[scale=0.4]{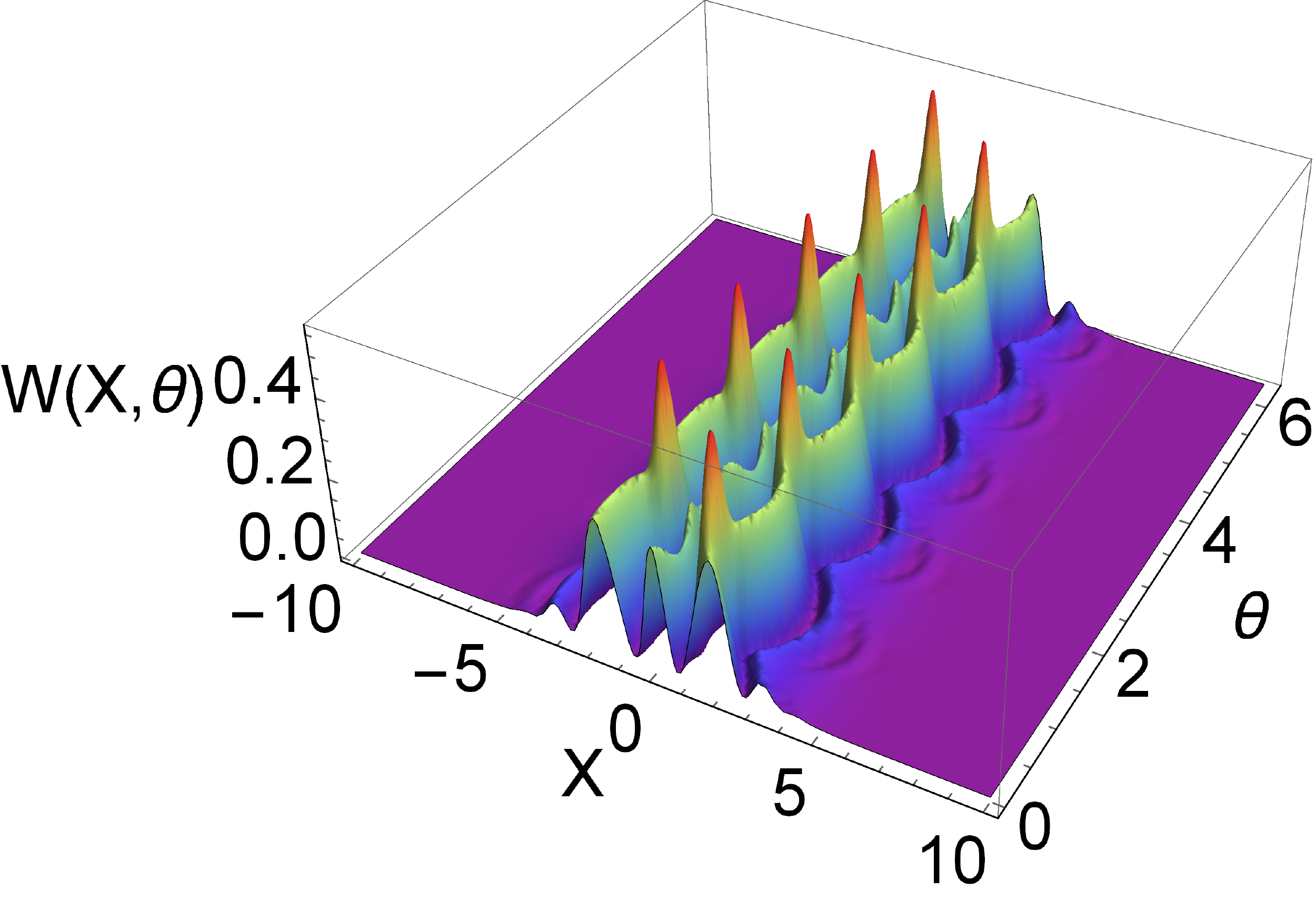} \
\includegraphics[scale=0.25]{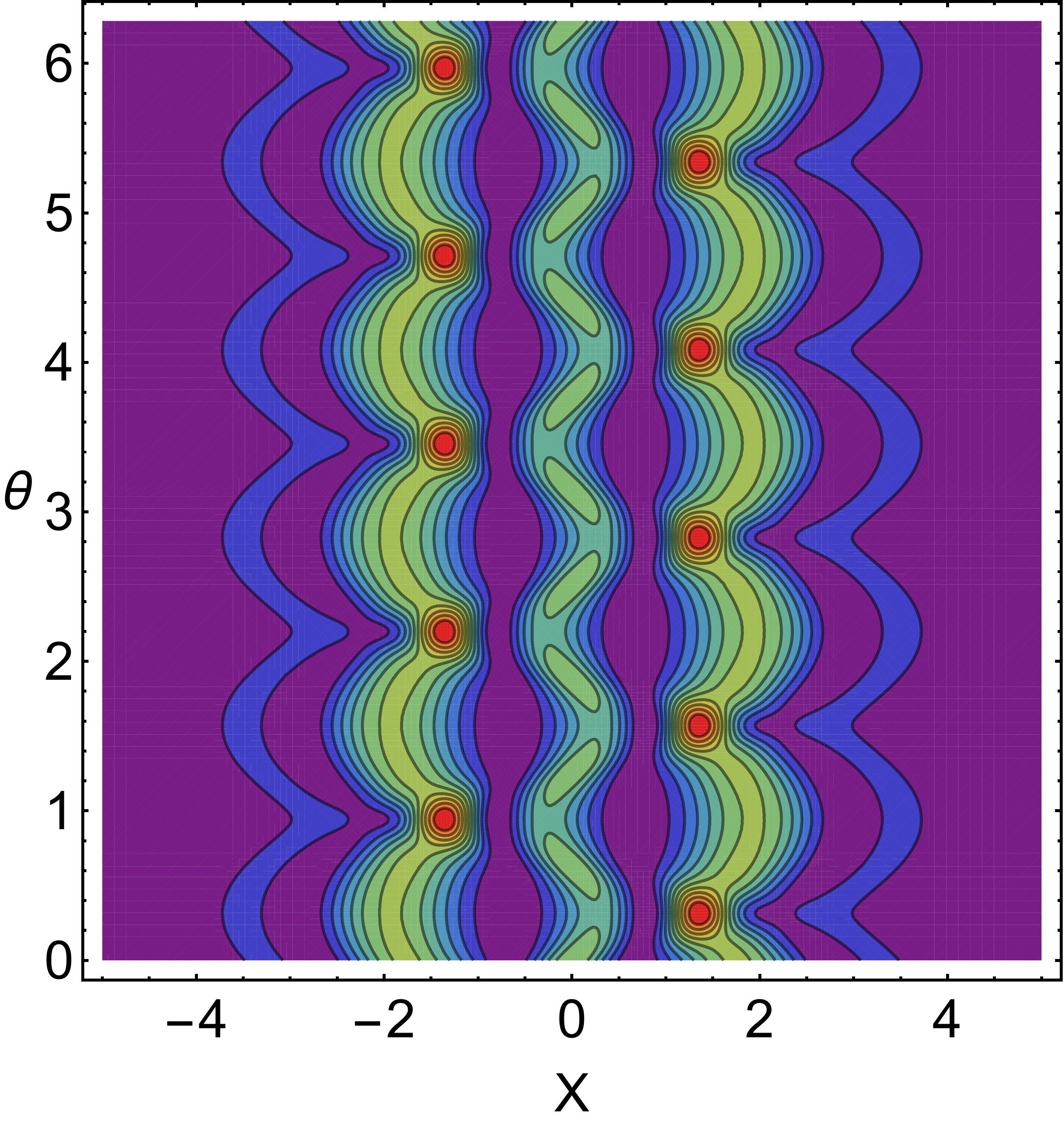}
\caption{Tomographic representation for the cyclic state of fifth order $C_4$
associated to the irreducible representation $\lambda=3$~(left) and its
contour plot~(right). Here, the parameters $a=2$, $b=2i$, and $s=1$ were
used.\label{cyc5}}
\end{figure}

In Figs.~\ref{cyc4} and \ref{cyc5}, the tomographic representation for the
cyclic states of fourth order ($\lambda=2$) and fifth order ($\lambda=3$) are
plotted in terms of the parameters $X$ and $\theta$. In these figures, we can
see that the probability of having a specific value of $X$ is split into
fringes, and the number of fringes depends on the order of the cyclic group.

In a similar way as the Wigner function, the tomogram associated to the
dihedral states can be obtained by the following formula:
\begin{equation}
w_\lambda(X,\theta)= \frac{N_\lambda^2}{2\pi \hbar} \sum_{r,r'=1}^n (w_{r,r'}
(X,\theta)+w_{r,r'}^\prime (X,\theta)+w_{r,r'}^{\prime \prime} (X,\theta)
+w_{r,r'}^{\prime \prime \prime} (X,\theta)),
\end{equation}
with the following integrals:
\begin{eqnarray}
w_{r,r'}^\prime (X,\theta)=\int_{-\infty}^\infty W^\prime_{r,r'}(X,P)\, dP,
\quad w_{r,r'}^{\prime \prime}(X,\theta)=\int_{-\infty}^\infty W^{\prime
\prime}_{r,r'}(X,P)\,  dP, \quad w_{r,r'}^{\prime \prime
\prime}(X,\theta)=\int_{-\infty}^\infty W^{\prime \prime \prime}_{r,r'}(X,P)
\, dP.\nonumber\\
\end{eqnarray}
Analogously to the Wigner function case mentioned above, the expressions for
$w_{r,r'}^\prime$, $w_{r,r'}^{\prime \prime}$, and $w_{r,r'}^{\prime \prime
\prime}$ can be directly calculated from Eq.~(\ref{tomrs}) by making the
proper substitutions.

\begin{figure}
\centering
\includegraphics[scale=0.4]{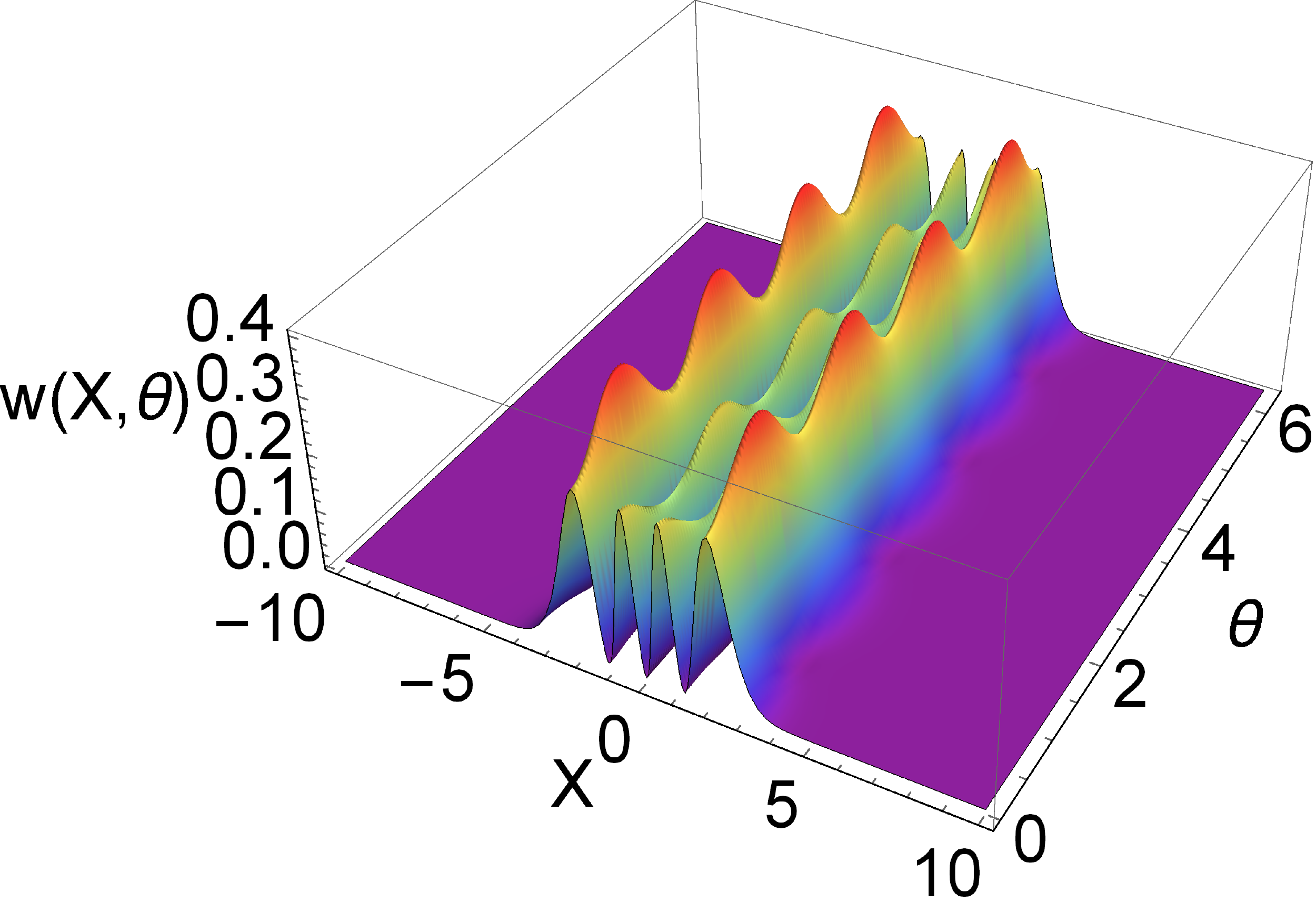} \
\includegraphics[scale=0.25]{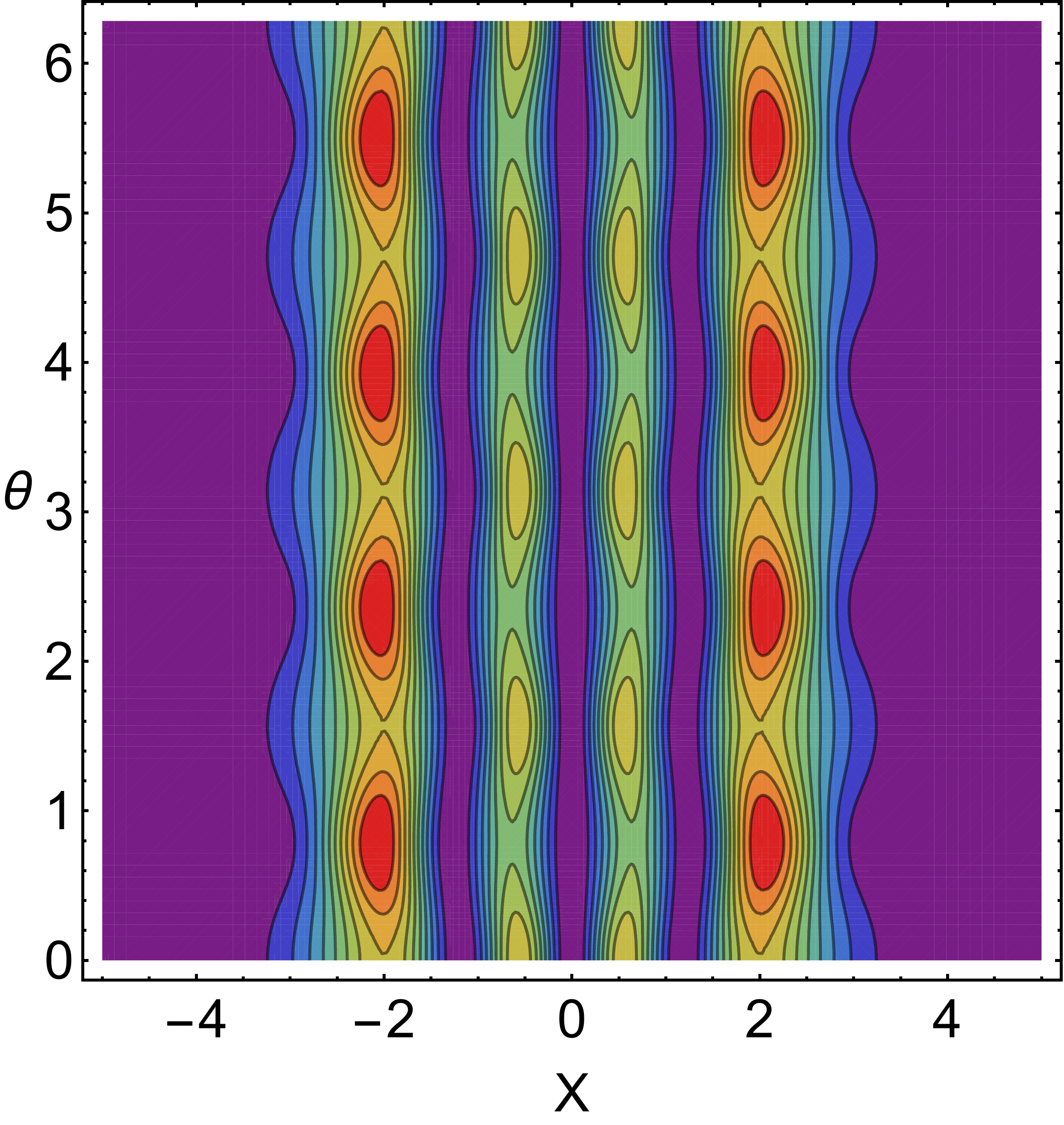}
\caption{ Tomogram for the dihedral state of fourth degree and irreducible
representation $\lambda=4$ as a function of $X$ and $\theta$~(left) and its
contour plot~(right). Here, the parameters $a=1$, $b=1/2$, and $s=1$ were
taken. \label{dih4}}
\end{figure}

\begin{figure}
\centering
\includegraphics[scale=0.4]{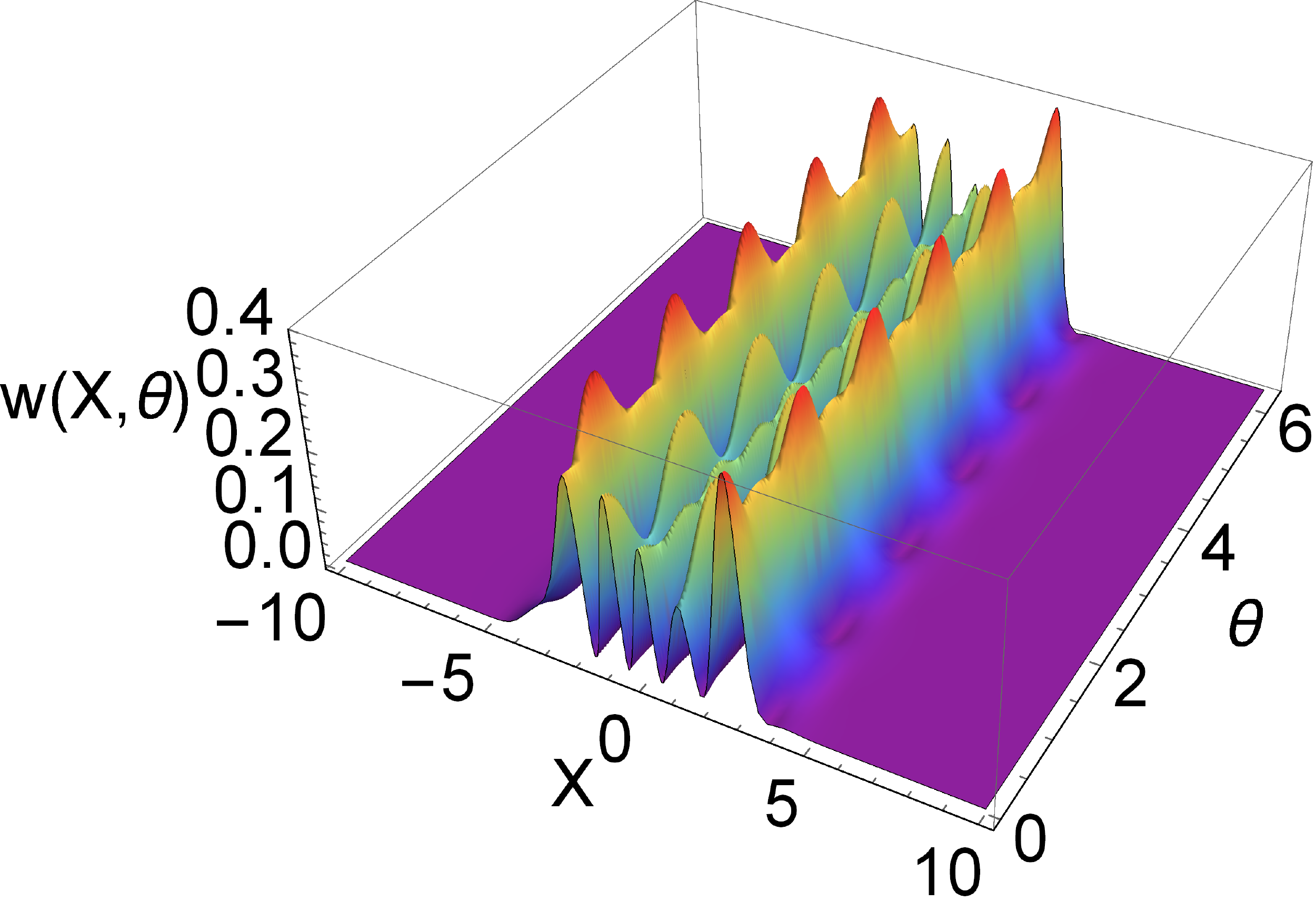} \
\includegraphics[scale=0.25]{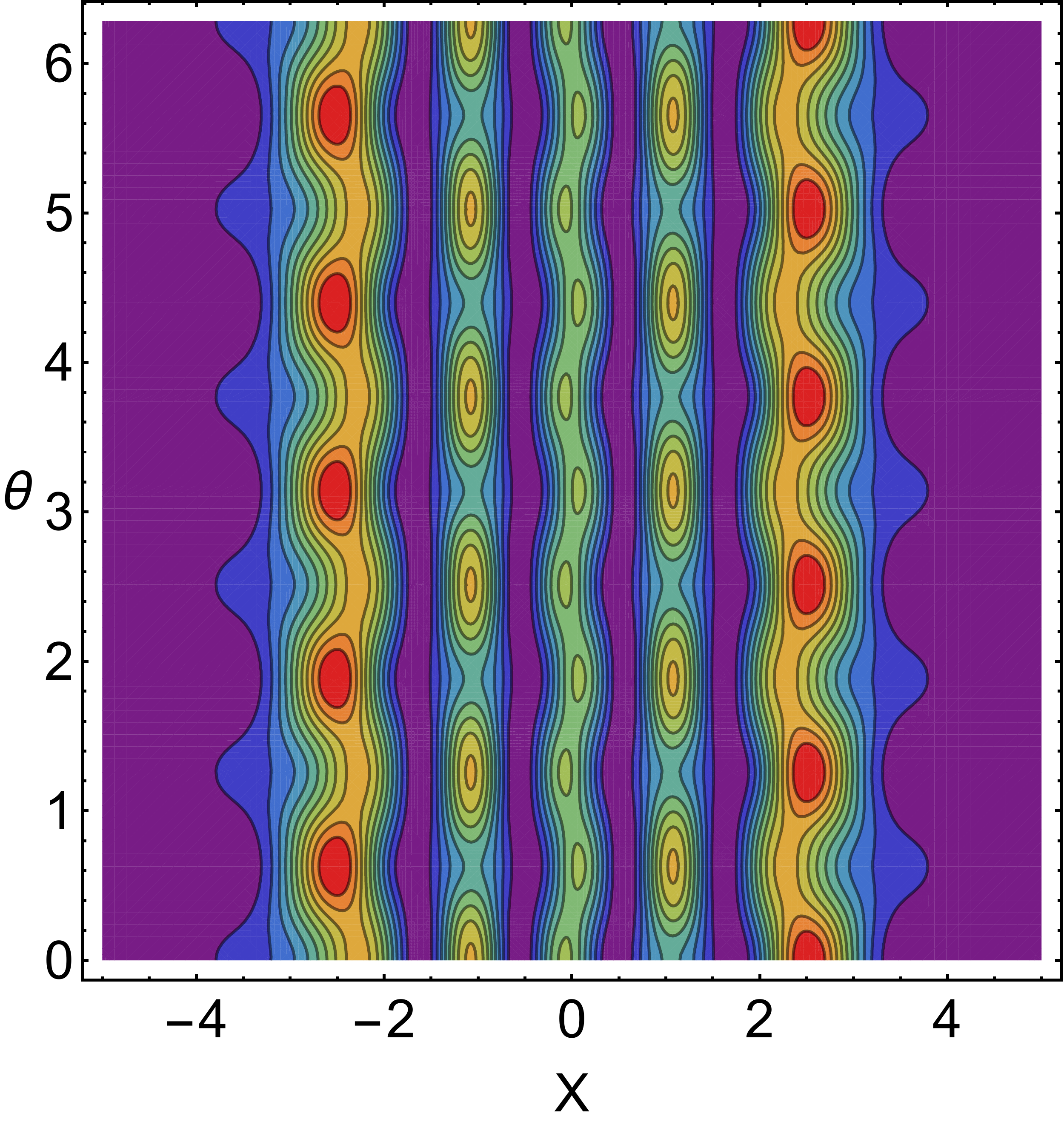}
\caption{Tomogram for the dihedral state of fifth degree and irreducible
representation $\lambda=5$ as a function of $X$ and $\theta$~(left) and its
contour plot~(right). Here, the parameters $a=1$, $b=1$, and $s=1$ were used.
\label{dih5}}
\end{figure}

Some examples of the resulting tomograms for the dihedral groups of fourth and
fifth order are shown in Figs.~\ref{dih4} and \ref{dih5}. One can see that, as in the cyclic case,
for those states, the probability of having a certain value for the
tomographic variable $X$ can be divided into different fringes, where the
probability of fringes depends on the degree of the dihedral group associated
to the symmetric states.

\section{Summary and concluding remarks}
We point out the main results of our work. The symmetric superposition of cyclic and dihedral states where obtained in the probabilistic representation of quantum mechanics. In this representation, as an alternative of
the standard description of quantum system states given by complex wave
functions or density operators acting on the vectors of Hilbert spaces, the
usual probability distributions, also known as tomograms, determine the
states~\cite{ManciniPLA,Olga-Entropy21}. We studied particular states of
systems with continuous variables like oscillators and considered Gaussian
state superpositions, since the probability distributions of such states are
either normal probability distributions or sums of the Gaussian terms, which
have not been discussed in the literature but are considered here as examples
of Schr\"odinger cat states.

We reviewed the tomographic probability representation of quantum states and
its properties. The parti\-cular representation for the superposition of
coherent states for the cyclic group $C_3$ was explicitly addressed. Then the
Wigner quasiprobability distribution associated to the cyclic and dihedral
groups for a superposition of generic states was presented and used
to obtain the tomographic probability representation of such states.

The obtained Wigner functions, for both cyclic and dihedral states,
demonstrate nonclassical behavior as they contain negative values for
different values of the position and momentum (or electromagnetic
quadratures). This is due to the interference between two different rotations
of the original Gaussian state, i.e., $\phi_j(x)$ and $\phi_{j'}(x)$ for
$j\neq j'$  in the definition of Wigner function.

In the tomographic probability representation, the probability distribution
for the parameter $X$ is different from zero in a series of fringes as a
function of the angle $\theta$. The number of fringes depends on the degree of
the cyclic or dihedral group, which defines a particular state.


\section*{Acknowledgements}
This work was partially supported by DGAPA-UNAM under Project IN101619.
Results of Sections 2 and 3 were obtained by V.I.M and J.A.L.S in the Russian
Quantum Center with the support from the Russian Science Foundation Grant No.
19-71-10091.


\begin{thebibliography}{99}

\bibitem{Schroedinger26}
E.  Schr\"odinger, Quantisierung als Eigenwertproblem (Zweite Mitteilung),
Ann. Phys.  {\bf 384} 361 and 489 (1926). doi:10.1002/andp.19263840404

\bibitem{Landau}
L. Landau, Das D\"ampfungsproblem in der Wellenmechanik, Z. Phys. {\bf 45} 430
(1927).

\bibitem{vonNeimann}
J. von Neumann, Wahrscheinlichkeitstheoretischer Aufbau der Quantenmechanik,
G\"ott. Nach. 245 (1927).
\newline
J. von Neumann, Mathematical foundations of Quantum Mechanics (Princeton
University Press, USA), 1955.

\bibitem{ManciniPLA}
S.~Mancini, V. I. Man'ko, and P.~Tombesi, Symplectic tomography as classical
approach to quantum systems, Phys. Lett. A \textbf{213(1)} 1 (1996).
doi:10.1016/0375-9601(96)00107-7

\bibitem{IbortPS150}
M.~Asorey, A.~Ibort, G.~Marmo, and F.~Ventriglia, Quantum tomography twenty
years later, Phys. Scr. \textbf{90(7)} 074031 (2015).
doi:10.1088/0031-8949/90/7/074031

\bibitem{Olga-Entropy21}
O. V. Man'ko and V. I. Man'ko, Probability representation of quantum states,
Entropy \textbf{23(5)} 549 (2021). doi:{10.3390/e23050549}

\bibitem{Glauber63}
R. J. Glauber, Coherent and incoherent states of the radiation field, Phys.
Rev. \textbf{131} 2766 (1963). doi{10.1103/PhysRev.131.2766}

\bibitem{Sudar63}
E. C. G. Sudarshan, Equivalence of semiclassical and quantum mechanical
descriptions of statistical light beams, Phys. Rev. Lett. \textbf{10} 277
(1963). doi:10.1103/PhysRevLett.10.277

\bibitem{Sanders-PR}
Barry C. Sanders, Entangled coherent states, Phys. Rev. A {\bf 45} 6811
(1992); Erratum, Phys. Rev. A {\bf 46} 2966 (1992).
doi:10.1103/PhysRevA.45.6811

\bibitem{Sanders-JPA}
B. C. Sanders, Review of entangled coherent states, J. Phys. A: Math. Theor.
{\bf 45(24)} 244002 (2012). doi:10.1088/1751-8113/45/24/244002

\bibitem{DodPhysica1974}
V. V. Dodonov, I. A. Malkin, and  V. I. Man'ko, Even and odd coherent states
and excitations of a singular oscillator,  Physica \textbf{72(3)} 597 (1974).
doi:10.1016/0031-8914(74)90215-8

\bibitem{Olga-SPIE}
S. V. Kuznetsov, A. V. Kyusev, O. V. Man'ko, and N. V. Chernega, Tomography
and satistical properties of superposition states for two--mode systems,
Bulletin of the Russian Academy of Science, Physics, Alerton Press Inc., Vol.
68, No 9, pp.~1239-1243 (2004) [in Russian].
https://www.elibrary.ru/item.asp?id=17641248 [Translatred into English by
Pleiades Publishing amd distributed by Springer (Springer Nature Switzerland
AG)].
\newline S. V. Kuznetsov, A. V. Kyusev, O. V. Man'ko,
Tomography and statistical properties of superposition states for two-mode
systems, Proc. SPIE, Vol. 5402, International Workshop on Quantum Optics 2003,
St.-Peterburg (2004) p.~314-327. doi:10.1117/12.562263

\bibitem{Adam}
P. \'Ad\'am, M. A. Man'ko, and V. I.~Man'ko, Even and odd Schr\"odinger cat
states in the probability representation of quantum mechanics, J. Russ. Laser
Res. {\bf 43} 1 (2022).

\bibitem{manko2}
V. V.  Dodonov and V. I. Man'ko, Proceedings of the Lebedev Physical Institute
{\bf 183} (Commack, NY: Nova Science) (1989).

\bibitem{manko3}
O. Casta\~nos, R. L\'opez-Pe\~na, and V. I. Man'ko, Crystallized
Schr\"{o}dinger cat states, J. Russ. Laser Res. {\bf 16} 477 (1995).
doi:10.1007/BF02581033

\bibitem{castanos}
O. Casta\~nos and J. A. L\'opez-Sald\'ivar, Dynamics of Schr\"{o}dinger cat
states, J. Phys.: Conf. Ser. {\bf 380} 012017 (2012).
doi:10.1088/1742-6596/380/1/012017

\bibitem{entan}
J. A. L\'opez-Sald\'ivar, A. Figueroa, O. Casta\~nos, R. L\'opez-Pe\~na, M. A.
Man'ko, and V. I. Manko, Evolution and entanglement of Gaussian states in the
parametric amplifier, J. Russ. Laser Res. {\bf 37} 23 (2016).
doi:10.1007/s10946-016-9543-2

\bibitem{generalstates}
J. A. L\'opez-Sald\'ivar, General superposition states associated to the
rotational and inversion symmetries in the phase space, Phys. Scri. {\bf 95}
065206 (2020). doi:10.1088/1402-4896/ab7

\bibitem{ourjoumtsev1}
A. Ourjoumtsev, R. Tualle-Brouri, J. Laurat, and P. Grangier, Generating
optical Schr\"odinger kittens for quantum information processing, Science {\bf
312} 83 (2006). doi:10.1126/science.1122858


\bibitem{wang}
B. Wang and L.-M. Duan, Engineering superpositions of coherent states in
coherent optical pulses through cavity-assisted interaction, Phys. Rev. A {\bf
72} 022320 (2005). doi:10.1103/PhysRevA.72.022320

\bibitem{hacker}
B. Hacker, S. Welte, S. Daiss, A. Shaukat, S. Ritter, L. Li, and G. Rempe,
Deterministic creation of entangled atom-light Schr\"{o}dinger-cat states,
Nature Photon. {\bf 13} 110 (2019). doi:10.1038/s41566-018-0339-5

\bibitem{ourjoumtsev2}
A. Ourjoumtsev, H. Jeong, R. Tualle-Brouri, and P. Grangier, Generation of
optical Schr\"dinger cats from photon number states, Nature {\bf 448} 784
(2007). doi:10.1038/nature06054

\bibitem{takahashi}
H. Takahashi, K. Wakui, S. Suzuki, M. Takeoka, K. Hayasaka, A. Furusawa, and
M. Sasaki, Generation of large-amplitude coherent-state superposition via
ancilla-assisted photon subtraction, Phys. Rev. Lett. {\bf 101} 233605 (2008).
doi:10.1103/PhysRevLett.101.233605

\bibitem{gerrits}
T. Gerrits, S. Glancy, T. S. Clement, B. Calkins, A. E. Lita, A. J. Miller, A.
L. Migdall, S. W. Nam, R. P. Mirin, and E. Knill, Generation of optical
coherent-state superpositions by number-resolved photon subtraction from the
squeezed vacuum, Phys. Rev. A {\bf 82} 031802(R) (2010).
doi:10.1103/PhysRevA.82.031802

\bibitem{neegaard}
J. S. Neergaard-Nielsen, B. Melholt Nielsen, C. Hettich, K. M\"almer, and E.
S. Polzik, Generation of a superposition of odd photon number states for
quantum information networks, Phys. Rev. Lett. {\bf 97} 083604  (2006).
doi:10.1103/PhysRevLett.97.083604

\bibitem{janszky}
J. Janszky, P. Domokos, and P. \'Ad\'am, Coherent states on a circle and
quantum interference, Phys. Rev. A {\bf 48} 2213 (1993).
doi:10.1103/PhysRevA.48.2213

\bibitem{domokos}
P. Domokos, P. \'Ad\'am, and J. Janszky, One-dimensional coherent-state
representation on a circle in phase space, Phys. Rev. A {\bf 50} 4293 (1994).
doi:10.1103/PhysRevA.50.4293

\bibitem{buzek}
V. Bu\~zek, A. Vidiella-Barranco, and P. L. Knight, Superpositions of coherent
states: Squeezing and dissipation, Phys. Rev. A {\bf 45} 6570 (1992).
doi:10.1103/PhysRevA.45.6570

\bibitem{szabo}
S. Szabo, P. \'Ad\'am, J. Janszky, and P. Domokos, Construction of quantum
states of the radiation field by discrete coherent-state superpositions, Phys.
Rev. A {\bf 53} 2698 (1996). doi:10.1103/PhysRevA.53.2698

\bibitem{gonzalez}
J. A. Gonz\'alez and M. A. del Olmo, Coherent states on the circle and
quantization, J. Phys. A {\bf 31} 8841 (1998). doi:10.1088/0305-4470/31/44/012

\bibitem{susskind}
L. Susskind and J. Glogower, Quantum-mechanical phase and time operator,
Physics Physique Fizika  {\bf 1} 49 (1964).
doi:10.1103/PhysicsPhysiqueFizika.1.49

\bibitem{nieto1}
M. M. Nieto, Quantum phase and quantum phase operators: some physics and some
history Phys. Scr. {\bf T48} 5 (1993). doi:10.1088/0031-8949/1993/T48/001

\bibitem{pegg1}
D. T. Pegg and S. M. Barnett, Unitary phase operator in quantum mechanics,
Europhys. Lett. {\bf 6} 483 (1988). doi:10.1209/0295-5075/6/6/002

\bibitem{pegg2}
D. T. Pegg and S. M. Barnett, Phase properties of the quantized single-mode
electromagnetic field, Phys. Rev. A {\bf 39} 1665 (1989).
doi:10.1103/PhysRevA.39.1665

\bibitem{calixto}
M. Calixto, J. Guerrero, and J. C. S\'anchez-Monreal, Sampling theorem and
discrete Fourier transform on the Riemann sphere, J. Fourier Anal. Appl. {\bf
14} 538 (2008). doi:10.1007/s00041-008-9027-z

\bibitem{calixto1}
M. Calixto, J. Guerrero, and J. C. S\'anchez-Monreal, Sampling heorem and
discrete Fourier transform on the hyperboloid, J. Fourier Anal. Appl. {\bf 17}
240 (2011). doi:10.1007/s00041-010-9142-5

\bibitem{cordero1}
O. Casta\~nos, S. Cordero, E. Nahmad-Achar, and R. L\'opez-Pe\~na, Generation
and dynamics of crystallised-type states of light within the Tavis--Cummings
model. In: Duarte S., Gazeau J.-P., Faci S., Micklitz T., Scherer R., and
Toppan F. (Eds.), Physical and Mathematical Aspects of Symmetries (Springer,
Cham, 2017). doi:10.1007/978-3-319-69164-0\_17

\bibitem{cordero2}
S. Cordero, E. Nahmad-Achar, O. Casta\~nos, and R. L\'opez-Pe\~na, Dynamic
generation of light states with discrete symmetries, Phys. Rev. A {\bf 97},
013808 (2018). doi:10.1103/PhysRevA.97.013808

\bibitem{MendesPLA}
V. I. Man'ko and R.~Vilela Mendes, Phys. Lett. A \textbf{263} 53 (1999).
doi:10.1016/S0375-9601(99)00688-X


\end{thebibliography}
\end{document}